\documentclass[a4paper]{jpconf}
\usepackage{graphicx}
\usepackage{cite}
\usepackage{amsmath}
\begin{document}
\title{Ternary $Z_2 \times Z_3$ graded algebras \\ and ternary Dirac equation}

\author{Richard Kerner}

\address{Laboratoire de Physique Th\'eorique de la Mati\`ere Condens\'ee (LPTMC), Univesit\'e Pierre et Marie Curie - CNRS UMR 7600
Tour 23-13, 5-\`eme \'etage, Bo\^{i}te Courrier 121, 4 Place Jussieu, 75005 Paris, FRANCE}

\ead{richard.kerner@upmc.fr}
\vskip 0.7cm

\begin{abstract}

The wave equation generalizing the Dirac operator to the $Z_3$-graded case is introduced, 
whose diagonalization leads to a sixth-order equation. It intertwines not only quark and anti-quark state
as well as the $u$ and $d$ quarks, but also the three colors, and is therefore invariant under
the product group $Z_2 \times Z_2 \times Z_3$. The solutions of this equation cannot propagate
 because their exponents always contain non-oscillating real damping factor. 
We show how certain cubic products can propagate nevertheless.
The model suggests the origin of the color $SU(3)$ symmetry and of the $SU(2) \times U(1)$ that arise
automatically in this model, leading to the full bosonic gauge sector of the Standard Model.

\end{abstract}
%
%\maketitle
%
\section{Introduction}

According to the present knowledge, the ultimate undivisible and undestructible
constituents of matter, called {\it atoms} by ancient Greeks, are in fact
the {\it quarks}, carrying fractional electric charges and baryonic numbers, 
two features that appear to be undestructible and conserved under any circumstances.

The notion of {\it Quarks} was introduced in elementary particle physics
in the early sixties, in now historical papers of M.Gell-Mann, Y.Ne'eman, S.Okubo, A.Zweig (\cite{GellMannNeeman}, \cite{Lipkin}, \cite{Okubo})
and others. Since the great success of unitary symmetries in the classification
of elementary particles that followed soon later, and  since the   
spectacular development of quantum chromodynamics (QCD), the fact that
{\it isolated single} quarks can not be observed becomes more mysterious than ever.

Taken into account that quarks evolve inside nucleons as almost point-like
entities, one may wonder how the notions of space and time still apply in these conditions ?
Perhaps in this case the Lorentz invariance can be derived from some more fundamental
{\it discrete} symmetries underlying the interactions between quarks ?

 If this is the case, then the symmetry $Z_3$ must play a fundamental role.
The carriers of elementary charges go by packs of three:
there are three families of quarks, and three types of leptons.

But the most striking feature is the {\it color charge} carried by quarks, and
subjected to the exact $SU(3)$ symmetry. The only observable states of quarks
are those which mix three colors in equal proportions, giving the so called {\it white}, or {\it colorless}, combinations.
The color symmetry is fundamental in strong interactions, whereas the electroweak interactions
ignore the color charge.

In Quantum Chromodynamics quarks are considered as fermions, endowed  with spin $\frac{1}{2}$.  Only  {\it three}
quarks or anti-quarks can coexist inside a fermionic baryon (respectively, anti-baryon), and a pair 
quark-antiquark can form a meson with integer spin.
 Besides, they must belong to different {\it colors}, also a three-valued set. There are two quarks in the first generation, 
$u$ and $d$ (``up" and ``down"), which may be considered as two states of a more general object,
just like proton and neutron in $SU(2)$ symmetry are two isospin components of a nucleon doublet.
In contrast with electrons which cannot occupy the same state with the same spins, but there can be exactly two
electrons in the same state with opposite spins, the possibility of coexistence is open for {\it two} quarks in the same $u$-state or 
$d$-state, but not three.

Numerous explanations of the impossibility of observing a pure state
of single isolated quark have been proposed. We can cite here the  
hypothesis that quarks are indeed {\it magnetic monopoles};
 or that they  are confined in "bags" whose nature was not specified, but    
whose surface tension was supposed to be too strong to be destroyed by
energies actually at our disposal. An attractive potential proportional
to the distance between the quarks has been introduced, too, explaining
why it is difficult to separate them from each other.

Among these hypotheses, the idea  that is at the base of the {\it dual resonance model}
promoted by J.Scherk \cite{Scherk1975}, supposed that quarks might be just 
{\it artefacts}, somewhat like the poles of a long magnet,
 being conceived as the ends of an open relativistic string. 

It has developed later into
an independent new theory of {\it strings} and  {\it superstrings}.
This hypothesis seems to be the most radical of all, since it simply
rejects the notion of quarks as primary objects, advocating instead the
view in which they are perceived as an artefact.

Another brand of thinking is represented  by the ideas trying to       
endow quarks with properties so unusual, that their observability
is enhanced by purely algebraic effects: either by mutually    
annihilating interferences of corresponding wave functions,  or some
strong statistical effects akin to the Fermi-Dirac statictics that
exclude the possibility of observing two identical fermions in the
same quantum state at once.  

With quarks, one should explain the contrary: namely, why they can be   
observed only by quark-antiquark pairs, or pure quark or antiquark  
triplets. Such theories are known as "algebraic confinement"
or "para-statistics", using the 
generalized versions of group theory known under the name of {\it quantum groups}
 acting on {\it quantum spaces} whose coordinates do
not commute, but satisfy instead more general binary relations of
the type $xy = q yx$, $q $ being a complex number \cite{WessMadore}.

%\begin{equation}
%xy = q \ \ yx
%\end{equation}
%with  $q $ being a complex number \cite{WessMadore}.
In what follows, we shall investigate the consequences of this type
of algebraic confinement hypotheses, stressing the fact that the $Z_3 $-grading,
 the  {\it ternary algebras} and {\it tri-linear forms} appear
as the most natural and necessary ingredients of these constructions.

\section{Fundamental discrete symmetries $Z_2$ and $Z_3$}

As underlined above, the fact that baryons are composed of three quarks displaying three different colors suggests that
permutation groups $S_3$ and its cyclic group $Z_3$ must play an important role in any theory of strong interactions,
along with the well established $Z_2$ symmetries; the charge conjugation, the space reflection and the time reversal.

The discrete symmetries should act on the Hilbert space of quantum states, which is a linear vector space over the
field of complex numbers. Let us briefly recall the elementary properties of complex representations of $S_3$ and $Z_3$
groups.

We shall denote the primitive third root of unity by $j = e^{2 \pi i/3}$.
The cyclic abelian subgroup $Z_3$ contains three elements corresponding to the three
cyclic permutations, which can be represented via multiplication by $j$, $j^2$ and $j^3 =1$ (the identity).
{\small
\begin{equation}
\begin{pmatrix}
ABC \cr ABC
\end{pmatrix}
\rightarrow {\bf 1}, \, \ \ \, 
\begin{pmatrix}
ABC \cr BCA
\end{pmatrix}
\rightarrow {\bf j}, \, \ \ \,
\begin{pmatrix}
ABC \cr CAB
\end{pmatrix}
\rightarrow {\bf j^2},
\label{permutationseven}
\end{equation} }

%{Symmetries} 

\begin{figure}[hbt]
\centering 
\includegraphics[width=3cm, height=3.5cm]{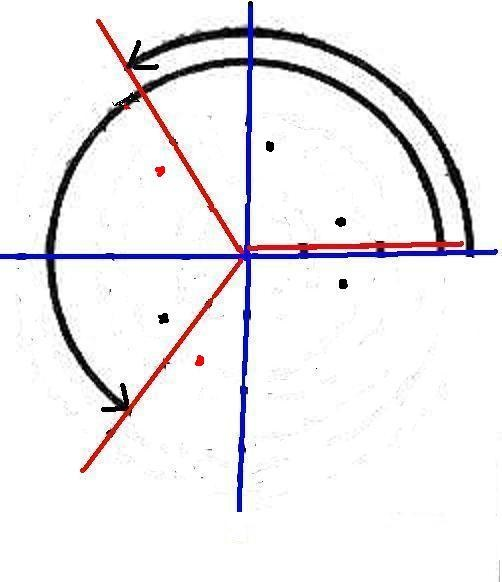} 
\hskip .5cm
\includegraphics[width=3cm, height=3.5cm]{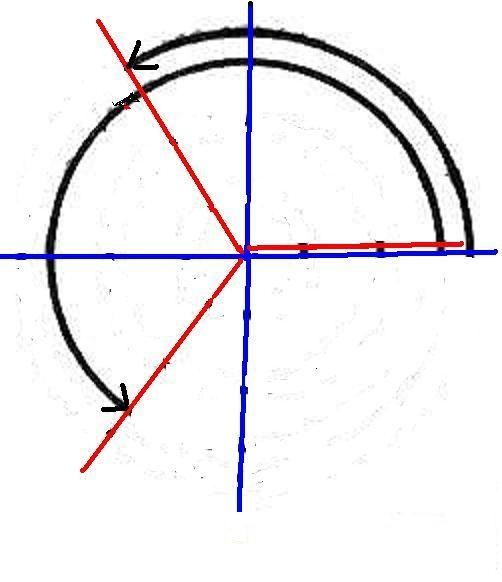} 
\caption{{\small The six $S_3$ symmetry transformations are: the identity, two rotations, one by $120^o$, another one  by $240^o$, and three
reflections, in the $x$-axis, in the $j$-axis and in the $j^2$-axis. The $Z_3$ subgroup consists only of the three rotations. } }
\label{fig:Rotations}
\end{figure}
%The six $S_3$ symmetry transformations contain the identity, two rotations, one by $120^o$, another one  by $240^o$, and three
%reflections, in the $x$-axis, in the $j$-axis and in the $j^2$-axis. The $Z_3$ subgroup contains only the three rotations.
%
It is important to note that if there are only two distinct states (e.g. $u$ and $d$), the full permutation group $S_3$
reduces itself to its abelian subgroup $Z_3$, because then even and odd permutations cannot be distinguished, e.g.
$uud \rightarrow udu \rightarrow duu$ are the only distinguishable permutations; one needs three different items to make
difference between even and odd permutations. The generalization of $Z_2$ grading to the $Z_3$ grading was proposed in 
\cite{Kerner1991}, \cite{Kerner1992}. 

\section{The $Z_6$ extension}

The presence of color degrees of freedom does not exclude the fundamental $Z_2$ symmetry between particles and anti-particles.
To take this into account we ought to consider the product group, which will result in the overall $Z_6 = Z_2 \times Z_3$ symmetry.
The cyclic group $Z_6$ is represented in the complex plane by all six powers of the non-trivial sixth root of unity, $q = e^{\frac{2 \pi i}{6}}.$
\begin{figure}[hbt]
\centering
\includegraphics[width=5cm, height=5cm]{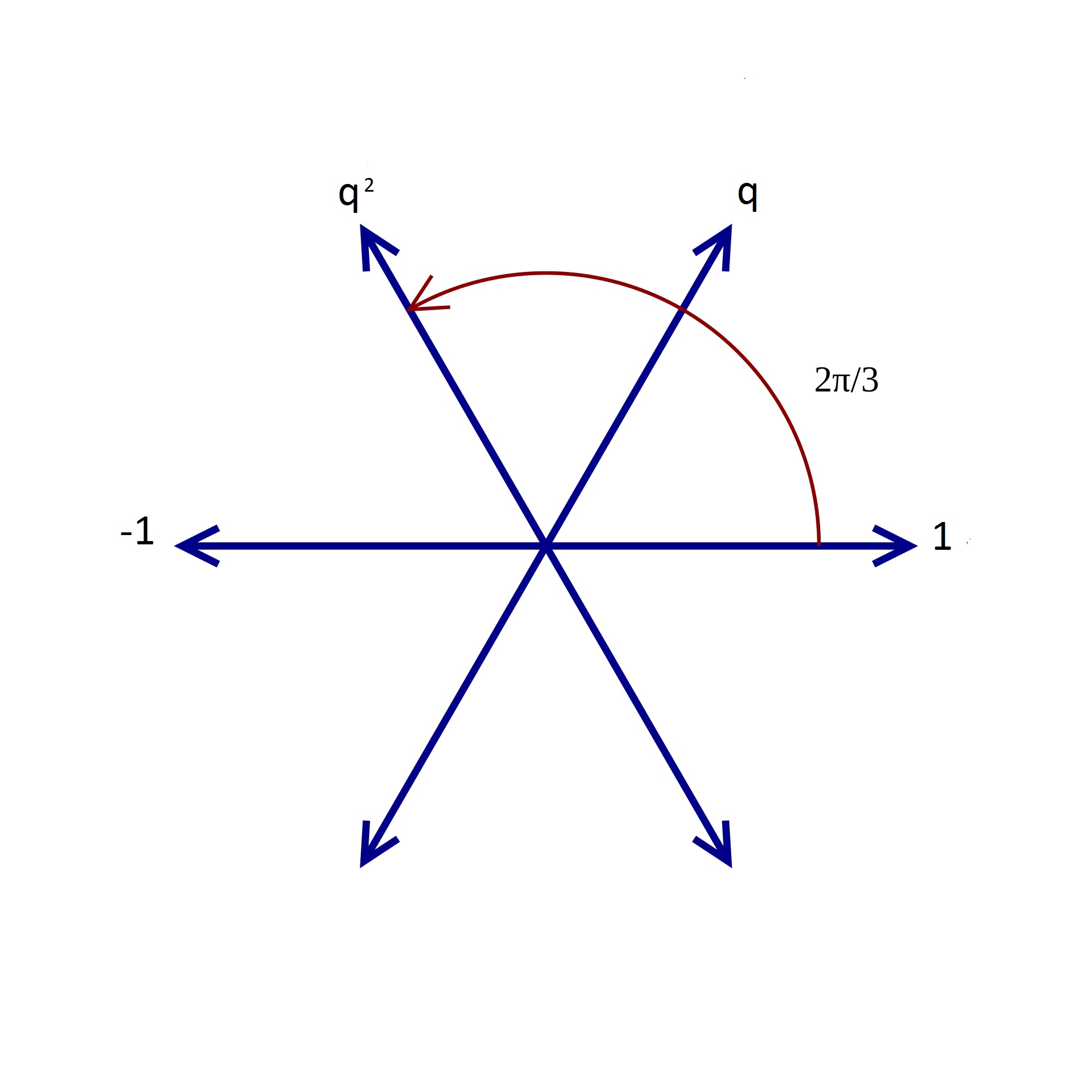}
\hskip 0.4cm
\includegraphics[width=4.5cm, height=4.5cm]{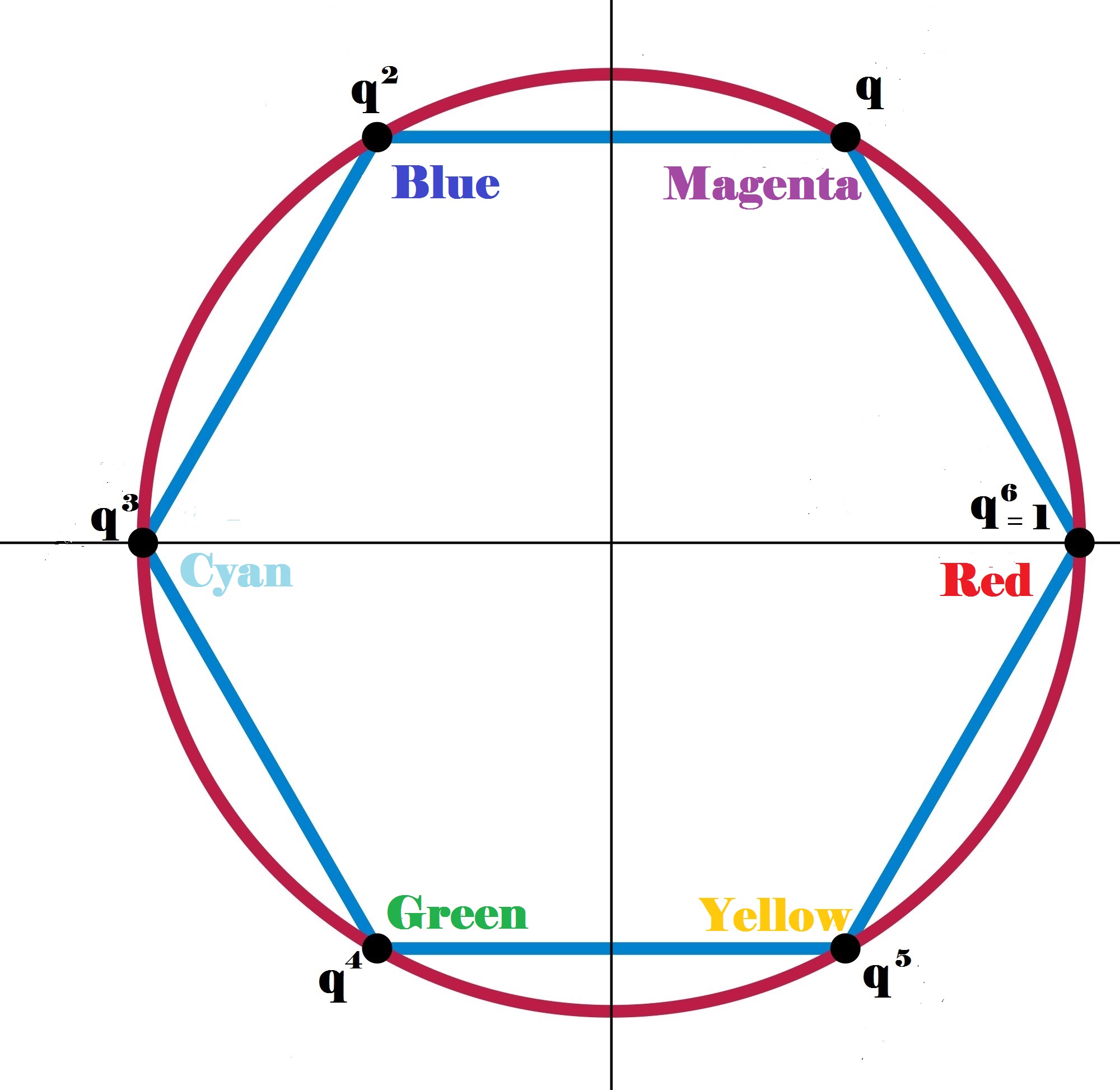}
\caption{\small{The six complex numbers $q^k$ can be put into correrspondence with three colors and three anti-colors.}}
\label{fig:CyclicZ6}
\end{figure}
The superselection rule according to which only observable superpositions of quark states must be
`white", or {\it colorless} can be implemented by simple algebraic rules of adding roots in $Z_6$ (see, e.g. \cite{AKL2015}).
If we identify zero in the complex plane as corresponding with ``white" combination of colors, we get
either two ``white" combinations of three colors:
\vskip 0.3cm
\indent
The three colors: $\; \; q^2 + q^4 + q^6 = j + j^2 + 1 = 0$.
\vskip 0.3cm
\indent
and the three anti-colors: \; \; \;  $q + q^3 +  q^5 = -j^2 - 1 - j = 0,$
\vskip 0.3cm
or the following three ``white" combinations of color with its anti-color:
\vskip 0.2cm
$q^0$ +  $q^3$ =  $1$  +  $(-1)$ = 0, \; \; \; $q^2$ + $q^5$ = $j$  +  $(-j)$ = 0, \; \; \; $q^4$ +  $q$ =  $j^2$  +  $(-j^2)$ = 0,

These sets of colors should be attributed to three quarks or three anti-quarks, or to the quark-antiquark pairs
according to the well known scheme: $uud$ for the proton, $udd$ for the neutron, ${\bar{u}}{\bar{u}}{\bar{d}}$ for 
the anti-proton and ${\bar{u}}{\bar{d}}{\bar{d}}$ for the anti-neutron. The three pi-mesons are identified with the
following quark-antiquark pairs: $u {\bar{d}}$ the $\pi^+$, ${\bar{u}} d$ the $\pi^{-}$ and $\frac{1}{\sqrt{2}} (u {\bar{u}} - d {\bar{d}})$
the $\pi^0$ meson respectively. 

The {\it gluons} are the gauge fields that mediate strong interactions between quarks; they provoke color exchange, but
in such a way that both initial and final states remain strictly ``white". In the language of second quantization all interactions 
are composed of creation and annihilation operators: let $B$ symbolize the creation of elementary ``blue" charge, and
let ${\bar{B}}$ denote the operator of its annihilation (equivalent with creation of its ``anti-color", (yellow in this particular case), etc.  
Basic interaction consists of replacing of one color by another, e.g. $B {\bar{G}}$ results in replacing green by blue. But such 
single action performed on a colorless combination will result in two blue charges and one red, which is not ``white" anymore. 
This is why the operator $B {\bar{G}}$ should be always accompanied by its hermitian conjugate $G {\bar{B}}$. This leads to the following
set of color interaction carriers:   

%However, the totally ``colorless" quadratic combinations (gluons) do not interact
%strongly with quarks. This means that the combination
%$$R {\bar{R}} + B {\bar{B}} + G {\bar{G}} = 0.$$
% does vanish, so that only {\it eight} linear combinations out of nine are independent.
%
%\subsection{Colors and the $Z-6$ group}

$$\frac{1}{\sqrt{2}} \; (R {\bar{B}} + B {\bar{R}}), 
\; \; \; \frac{1}{i \sqrt{2}} (R {\bar{B}} - B {\bar{R}}), \; \; \; 
\frac{1}{\sqrt{2}} \; (R {\bar{G}} + G {\bar{R}}), $$
$$ \frac{1}{i \sqrt{2}} (R {\bar{G}} - G {\bar{R}}), \; \; \; 
\frac{1}{\sqrt{2}} \; (B {\bar{G}} + G {\bar{B}}), 
\; \; \; \frac{1}{i \sqrt{2}} (B {\bar{G}} - G {\bar{B}}),$$
$$\frac{1}{\sqrt{2}} \; (R {\bar{R}} - B {\bar{B}}), 
\; \; \; \frac{1}{ \sqrt{6}} (R {\bar{R}} +  B {\bar{B}} - 2 G {\bar{G}}),$$

Acting on a column with three entries, $R, B$ and $G$, these combinations form the basis of eight traceless Gell-Mann $3 \times 3$ matrices.

The totally ``colorless" quadratic combination does not interact
strongly with quarks - it corresponds to the unit matrix, i.e. does not change the color content of the initial configuration.
Therefore, the combination
$$R {\bar{R}} + B {\bar{B}} + G {\bar{G}} = 0.$$
is not taken into account, so that only {\it eight} linear combinations out of nine are independent, forming the basis of the
$SU(3)$ algebra generators.

\section{Ternary Dirac equation for colors}

Before introducing the generalized Dirac equation incorporating the color degrees of freedom, let
us remind how the spin degrees of freedom were first introduced by W. Pauli. Historically, as early as in 1925,
Pauli arrived at the conclusion that in order to understand electron's energy levels in atoms, a fourth quantum number is needed,
subjected to the {\it exclusion principle}, by now bearing Pauli's name. The identification of this dychotomic
quantum number with electron's spin was made by G. Uhlenbeck and S. Goudsmit, \cite{UhlenbeckGoudsmit} who explored experimentally 
the anomalous Zeeman effect. In $1927$ Pauli \cite{Pauli} introduced the equation (also bearing his name today) describing the interaction
of electron's spin with external magnetic field. But the way he arrived to this equation and how he missed the relativistic
 equation for the electron (introduced almost immediately after by P.A.M. Dirac) is very inspiring indeed. 

The inclusion of spin variable, subjected to Pauli's exclusion principle, into a Schroedinger-like equation can be done
 by replacing the usual complex wave function by a column vector containing two complex components. The energy, momentum
and mass operators should be represented by $2 \times 2$ matrices. The simplest linear equation considered by Pauli at first
had the following form:
\begin{equation}
E \; {\mbox{l\hspace{-0.55em}1}}_2 \; \psi = mc^2 \;  {\mbox{l\hspace{-0.55em}1}}_2 \; \psi + c \; {\boldsymbol{\sigma}}\cdot {\bf p} \; \psi,
\label{Paulifirst}
\end{equation}  
where $\psi$ stays now for the two-component {\it Pauli spinor} $\begin{pmatrix} \psi^1 \cr \psi^2 \end{pmatrix} $, the
$3$-dimensional momentum vector ${\bf p}$ is scalarly multiplied by ${\boldsymbol{\sigma}}$ representing the three hermitian
traceless Pauli's matrices ${\boldsymbol{\sigma}} = [ \sigma_x, \; \sigma_y, \; \sigma_z ]$ , and ${\mbox{l\hspace{-0.55em}1}}_2$
stays for the $2 \times 2$ unit matrix.
But this equation fails to satisfy the Lorentz invariance criterion: it suffices to take the square of the energy operator to
discover that (\ref{Paulifirst}) leads to the following quadratic relation
\begin{equation}
E^2 = m^2 c^4 + 2 m c^3 {\boldsymbol{\sigma}}\cdot {\bf p} + c^2 \; {\bf p}^2
\label{Epwrong}
\end{equation}
instead of the desired Lorentz-invariant relation $E^2 = m^2 c^4 + c^2 {\bf p}^2$.
At this stage the Lorentz invariance could be recovered by introducing another Pauli spinor entangled with the first one
via equations similar with (\ref{Paulifirst}), but with a negative mass term for the second Pauli spinor:
$$ E \; {\mbox{l\hspace{-0.55em}1}}_2 \; \psi_{+} = mc^2 \; {\mbox{l\hspace{-0.55em}1}}_2 \; \psi_{+} 
+ c \; {\boldsymbol{\sigma}}\cdot {\bf p} \; \psi_{-},$$
\begin{equation}
E \; {\mbox{l\hspace{-0.55em}1}}_2 \; \psi_{-} = - mc^2 \; {\mbox{l\hspace{-0.55em}1}}_2 \; \psi_{-} 
+ c \; {\boldsymbol{\sigma}}\cdot {\bf p} \; \psi_{+},
\label{Paulisecond}
\end{equation}
where $\psi_{+} = \begin{pmatrix} \psi_{+}^1 \cr \psi_{+}^2 \end{pmatrix}, \; \; \; \psi_{-} = \begin{pmatrix} \psi_{-}^1 \cr \psi_{-}^2 \end{pmatrix}$.
It is easy to see now that by simple iteration we get the right relation satisfied by simultaneously by both components:
$$E^2 \psi_{+} - c^2 {\bf p}^2 \; \psi_{+} = m^2 c^4 \psi_{+}, \; \; \; E^2 \psi_{-} - c^2 {\bf p}^2 \; \psi_{-} = m^2 c^4 \psi_{-}.$$
The four equations (\ref{Paulisecond}) are just one of the representations of the equation of the electron discovered shortly after by
Dirac, but in a totally different manner, derived as a ``square root" of the Klein-Gordon equation; but at the moment the idea
of introducing a negative mass seemed physically unacceptable. This is why Pauli opted for a non-relativistic equation for the electron
in the magnetic field, 
\begin{equation}
E \; \psi = \left[ \frac{1}{2m} ({\boldsymbol{\sigma}} \cdot ({\bf p} - e {\bf A}))^2 + e V(x) \right]  \psi.
= \left[ \frac{1}{2m} ({\bf p} - e {\bf A})^2 + e {\boldsymbol{\sigma}} \cdot {\bf B} + e V(x) \right]  \psi.
\label{Pauliequation}
\end{equation}
Later on it turned out that the Pauli equation (\ref{Pauliequation}) is the non-relativistic limit
of the Dirac equation. 

The two equations  (\ref{Paulisecond}) can be re-written using a matrix notation:
\begin{equation}
\begin{pmatrix} E & 0 \cr 0 & E \end{pmatrix} \, \begin{pmatrix} \psi_{+} \cr \psi_{-} \end{pmatrix}
= \begin{pmatrix} mc^2 & 0 \cr 0 & - mc^2 \end{pmatrix} \, \begin{pmatrix} \psi_{+} \cr \psi_{-} \end{pmatrix}
+ \begin{pmatrix} 0 & c \, {\boldsymbol{\sigma}} {\bf p} \cr {\boldsymbol{\sigma}} {\bf p} & 0 \end{pmatrix}
\; \begin{pmatrix} \psi_{+} \cr \psi_{-} \end{pmatrix},
\label{DiracPauli}
\end{equation}
where the entries in the energy operator and the mass matrix are in fact $2 \times 2$ identity
matrices, as well as the $\sigma$-matrices appearing in the last matrix, so that in reality
the above equation represents the $4 \times 4$ Dirac equation, only in a different basis \cite{Dirac}.

The system of linear equations (\ref{DiracPauli}) displays two important discrete $Z_2$ symmetries: the space reflection
consisting in simultaneous change of the direction of spin and momentum, ${\boldsymbol{\sigma}} \rightarrow - {\boldsymbol{\sigma}}, 
{\bf p} \rightarrow - {\bf p}$, and the particle-antiparticle symmetry realized by the transfromation $m \rightarrow - m, \; \psi_{+} \rightarrow \psi_{-},
\; \; \psi_{-} \rightarrow \psi_{+}$. Our next aim is to extend the $Z_2 \times Z_2$ symmetry by including the $Z_3$ group  which
will mix not only the two spin states and particles with anti-particles, but also the three colors. 

%\subsection{Inclusion of color degrees of freedom} 

Now we want to describe three different two-component fields (which can be incidentally given
the names of three colors, the ``red" one $\varphi_{+}$, the ``blue" one $\chi_{+}$, and the
``green" one $\psi_{+}$); more explicitly,

\begin{equation}
\varphi_{+} = \begin{pmatrix} \varphi_{+}^1 \cr \varphi_{+}^2 \end{pmatrix}, \;   
\chi_{+} = \begin{pmatrix} \chi_{+}^1 \cr \chi_{+}^2 \end{pmatrix}, \;   
\psi_{+} = \begin{pmatrix} \psi_{+}^1 \cr \psi_{+}^2 \end{pmatrix},
\end{equation}

We follow the minimal scheme  taking into account the existence of spin by using only Pauli spinors
on which the $3$-momentum operator acts through the scalar product ${\boldsymbol{\sigma}}\cdot {\bf p}$.
In order to satisfy the required existence of anti-particles, we should also introduce three
``anti-colors", denoted by a ``minus" underscript, corresponding to the opposite colors:
``cyan" for $\varphi_{-}$, ``yellow" for $\chi_{-}$ and ``magenta" for $\psi_{-}$; here, too,
we have to do with two-component columns:
$$
\varphi_{-} = \begin{pmatrix} \varphi_{-}^1 \cr \varphi_{-}^2 \end{pmatrix},  \; \;    
\chi_{-} = \begin{pmatrix} \chi_{-}^1 \cr \chi_{-}^2 \end{pmatrix}, \; \;  
\psi_{-} = \begin{pmatrix} \psi_{-}^1 \cr \psi_{-}^2 \end{pmatrix},$$
 all in all {\it twelve} components. A somewhat similar construction, but with three Dirac spinors, can be
found in \cite{Sogami2013}.

%This reflects the overall $Z_2 \times Z_2 \times Z_3$
%symmetry: one  $Z_2$ group corresponding to the spin-like dichotomic degree of freedom,
%describing two  (exclusively) accessible states; the second $Z_2$ required in  order to account for the 
%particle-anti-particle symmetry, and the $Z_3$ group corresponding to color symmetry.

 The ``colors" should satisfy first order equations conceived in such a way that neither
can propagate by itself, just like in the case of ${\bf E}$ 
and ${\bf B}$ components of Maxwell's tensor in electrodynamics, or the couple of two-component Pauli spinors 
which cannot propagate alone, but constitute one single entity, the four-component 
Dirac spinor. 

 This leaves little space for the choice of the system of intertwined equations; here
is the ternary generalization of Dirac's equation, intertwining not only particles with antiparticles,
but also the three ``colors", in such a way that the entire system becomes invariant under
the action of the $Z_2 \times Z_2 \times Z_3$ group. 

%\subsection{Dirac's equation with color mixing} 
 
The set of linear equations for three Pauli spinors endowed with colors, and another three Pauli spinors
corresponding to their anti-particles endowed with "anti-colors" involves altogether twelve complex functions.
The twelve components could describe three independent Dirac particles, but here they will be intertwined in
a particular manner, mixing together not only spin states and particle-antiârticle states, but also the three colors.

We shall follow the logic that led from Pauli's to Dirac's equation extending it to the colors acted upon
by the $Z_3$-group. In the expression for the energy operator (i.e. the Hamiltonian), mass terms is positive when
acting on particles, and acquires negative sign acting on anti-particles, i.e. it changes sign while intertwining
particle-antiparticle components. We shall also assume that the mass term acquires the factor $j$ when we switch
from the red component $\varphi$ to the blue component $\xi$, and $j^2$ for the green component $\psi$.
The momentum operator will be non-diagonal, as in the Dirac equation, systematically intertwining not only
particles with antiparticles, but also colors with anti-colors.  

The system that satisfies all these assumptions is as follows:

$$E \; \varphi_{+} = mc^2 \, \varphi_{+} + c \; {\boldsymbol{\sigma}} \cdot {\bf p} \, \chi_{-}$$
$$E \; \chi_{-} = - j \; mc^2 \, \chi_{-} + c \; {\boldsymbol{\sigma}} \cdot {\bf p} \, \psi_{+}$$
$$E \; \psi_{+} = j^2 \;  mc^2 \, \psi_{+} + c \; {\boldsymbol{\sigma}} \cdot {\bf p} \, \varphi_{-}$$
$$E \; \varphi_{-} = - mc^2 \, \varphi_{-} + c \; {\boldsymbol{\sigma}} \cdot {\bf p} \, \chi_{+}$$
$$E \; \chi_{+} = j \; mc^2 \, \chi_{+} + c \; {\boldsymbol{\sigma}} \cdot {\bf p} \, \psi_{-}$$
\begin{equation}
E \; \psi_{-} = -j^2 \;mc^2 \, \varphi_{+} + c \; {\boldsymbol{\sigma}} \cdot {\bf p} \, \varphi_{+}
\label{systemsix}
\end{equation}
where 
$$ \varphi_{+} = \begin{pmatrix} \varphi_{+}^1 \cr \varphi_{+}^2 \end{pmatrix}, \; \; 
\varphi_{-} = \begin{pmatrix} \varphi_{-}^1 \cr \varphi_{-}^2 \end{pmatrix}, \; \; 
\chi_{+} = \begin{pmatrix} \chi_{+}^1 \cr \chi_{+}^2 \end{pmatrix}, \; \; \; 
\chi_{-} = \begin{pmatrix} \chi_{-}^1 \cr \chi_{-}^2 \end{pmatrix}, \; \; 
\psi_{+} = \begin{pmatrix} \psi_{+}^1 \cr \psi_{+}^2 \end{pmatrix}, \; \; 
\psi_{-} = \begin{pmatrix} \psi_{-}^1 \cr \psi_{-}^2 \end{pmatrix},$$
on which Pauli sigma-matrices act in a natural way.

On the right-hand side, the mass terms form a diagonal matrix whose entries follow an 
ordered row of powers of the sixth root of unity $q = e^{\frac{2 \pi i}{6}}$. Indeed, we have 
$$m = q^6  m, \; \;  -j m = q^5  m, \; \; j^2 m = q^4  m, $$
$$- m = q^3  m, \; \; j m = q^2  m, \; \; - j^2 m = q  m.  $$
Let us start the diagonalisation of our system by deriving two third-order equations 
relating between them the $\varphi_{+}$ 
and $\varphi_{-}$ fields. By iterating the $E$ operator three times, we get the following equation:

$$ E^3 \; \varphi_{+} = m^3 c^6 \; \varphi_{+} - 2 j \; m^2 c^5 \, 
{\boldsymbol{\sigma}} \cdot {\bf p} \, \chi_{-} 
- 2 j \; m c^3 \, {\boldsymbol{\sigma}} \cdot {\bf p} \, \psi_{+} +  
\mid {\bf p} \mid^2 \; {\boldsymbol{\sigma}} \cdot {\bf p} \, \varphi_{-} $$

As one can see, at the third iteration diagonalisation is not yet achieved because of the presence,
besides the fields $\varphi_{+}$ and $\varphi_{-}$, of two other fields, namely $\psi_{+}$ and $\chi_{-}$.

Similar third order equations are produced when we start the iteration from any of the five 
remaining components; in all cases, they contain four terms mixing other components. 
The diagonalization of the system is achieved only at the sixth iteration.

The final result is extremely simple: all the components satisfy the same sixth-order equation, 

% Here we arrive at the following
%sixth-order equation satisfied simultaneously by $\varphi_{+}$ and $\varphi_{-}$:

$$E^6 \; \varphi_{+} = m^6 c^{12} \; \varphi_{+} + c^6 \mid {\bf p} \mid^6 \; \varphi_{+},$$
\begin{equation}
E^6 \; \varphi_{-} = m^6 c^{12} \; \varphi_{-} + c^6 \mid {\bf p} \mid^6 \; \varphi_{-}.
\label{E6varphi}
\end{equation}
and similarly for all other components.

The energy operator is obviously diagonal, and its action on the spinor-valued column-vector 
can be represented as a  $6 \times 6$ operator valued unit matrix.
The mass operator is diagonal, too, but its elements represent all powers of the sixth root of
unity $q$, which are  $q = - j^2, \; q^2 = j, \; q^3 = -1, \; q^2 = j^2, \; q^5 = - j$ and $q^6 =1$.

Finally, the momentum operator is proportional to a {\it circulant matrix}
 which mixes up all the components of the column vector.

In the basis in which the original system (\ref{systemsix}) was proposed, 
the matrix operators can be expressed as follows:
$$M = \begin{pmatrix} m & 0 & 0 & 0 & 0 & 0 \cr 0 & -m & 0 & 0 & 0 & 0  \cr
0 & 0 & j m & 0 & 0 & 0 \cr 0 & 0 & 0 & - j m & 0 & 0 \cr 0 & 0 & 0 & 0 & j^2 m & 0 \cr
0 & 0 & 0 & 0 & 0 & -j^2 m \end{pmatrix}, \; \; \; \;   
P = \begin{pmatrix} 0 & 0 & 0 & {\boldsymbol{\sigma}} \cdot {\bf p} & 0 & 0 \cr 
0 & 0 & {\boldsymbol{\sigma}} \cdot {\bf p} & 0 & 0 & 0 \cr
0 & 0 & 0 & 0 & 0 & {\boldsymbol{\sigma}} \cdot {\bf p} \cr 
0 & 0 & 0 & 0 & {\boldsymbol{\sigma}} \cdot {\bf p} & 0 \cr 
0 & {\boldsymbol{\sigma}} \cdot {\bf p} & 0 & 0 & 0 & 0 \cr
{\boldsymbol{\sigma}} \cdot {\bf p} & 0 & 0 & 0 & 0 & 0 \end{pmatrix}$$

In fact, the dimension of the two matrices $M$ and $P$  displayed  above is $12 \times 12$: 
all the entries in the first one are proportional to the 
$2 \times 2$ identity matrix, so that in the definition one should read 

$\begin{pmatrix} m & 0 \cr 0 & m \end{pmatrix}$ instead of $m$, 
$\begin{pmatrix} j m & 0 \cr 0 & j m \end{pmatrix}$ instead of $j \; m$, etc.

The entries in the second matrix $P$ contain $2 \times 2$ Pauli's sigma-matrices,
so that $P$ is also a $12 \times 12$ matrix. The energy operator $E$ is proportional to
the $12 \times 12$ identity matrix. 

\section{Ternary Clifford Algebra}

Using a more rigorous mathematical language the three operators can be expressed 
in terms of tensor products of matrices of lower dimensions. Let us introduce two following $3 \times 3$ matrices:
\begin{equation}
B = \begin{pmatrix} 1 & 0 & 0 \cr 0 & j & 0 \cr 0 & 0 & j^2 \end{pmatrix} \; \; {\rm and} \; \; 
Q_3 = \begin{pmatrix} 0 & 1 & 0 \cr 0 & 0 & 1 \cr 1 & 0 & 0 \end{pmatrix}
\label{BQmatrices}
\end{equation}
Then the $12 \times 12$ matrices $M$ and $P$ can be represented as the following  tensor products:
\begin{equation}
M = m \; B \otimes \sigma_3 \otimes {\mbox{l\hspace{-0.55em}1}}_2, \; \; \; \; 
P =  Q_3 \otimes \sigma_1 \otimes ({\boldsymbol{\sigma}} \cdot {\bf p}) 
\label{MPtensor1}
\end{equation}
with as usual, ${\mbox{l\hspace{-0.55em}1}}_2, \; \sigma_1 $ and $\sigma_3$ denote the well known $2 \times 2$  Pauli's matrices
$$ {\mbox{l\hspace{-0.55em}1}}_2 = \begin{pmatrix} 1 & 0 \cr 0 & 1 \end{pmatrix}, \; \; \; 
\sigma_1 = \begin{pmatrix} 0 & 1 \cr 1 & 0 \end{pmatrix}, \; \; \; 
\sigma_3 = \begin{pmatrix} 1 & 0 \cr 0 & -1 \end{pmatrix}. $$ 

The matrices $Q_a$ and $Q^{\dagger}_b$ span a very interesting ternary algebra. They were considered by Sylvester and Cayley already in
the XIX-th century \cite{Sylvester}. Out of three independent $Z_3$-graded ternary
combinations, only one leads to a non-vanishing result. One can check without much effort that both $j$ and $j^2$ skew 
ternary commutators do vanish:
$$\{ Q_1, Q_2, Q_3 \}_j = Q_1 Q_2 Q_3 + j Q_2 Q_3 Q_1 + j^2 Q_3 Q_1 Q_2 = 0, $$
$$\{ Q_1, Q_2, Q_3 \}_{j^2} = Q_1 Q_2 Q_3 + j^2 Q_2 Q_3 Q_1 + j Q_3 Q_1 Q_2 = 0,$$
and similarly for the odd permutation, $Q_2 Q_1 Q_3$.

On the contrary, the totally symmetric combination does not vanish; it is proportional 
to the $3 \times 3$ identity matrix $ {\mbox{l\hspace{-0.55em}1}} $:
\begin{equation}
 Q_a Q_b Q_c + Q_b Q_c Q_a + Q_c Q_a Q_b = \eta_{abc} \, {\mbox{l\hspace{-0.55em}1}}, \; \; \; a,b,... = 1,2,3.
\label{anticom}
\end{equation}
with $\eta_{abc}$ given by the following non-zero components:
$$
\eta_{111} = \eta_{222} = \eta_{333} = 1, \; \; \; \eta_{123} = \eta_{231} = \eta_{312} = 1, $$
\begin{equation} 
\eta_{213} = \eta_{321} = \eta_{132} = j^2.
\label{defeta}
\end{equation}
all other components vanishing. This relation may serve as the definition of {\it ternary Clifford algebra}.

Another set of three matrices is formed by the hermitian conjugates of $Q_a$, which coincide, with the {\it squares} of corresponding $Q_a$'s.
It is easy to check that one has
\begin{equation}
Q^2_a = Q_a^{\dagger}
\label{defqbar}
\end{equation}
The set of three conjugate matrices $Q^{\dagger}_a$ satisfy identities conjugate to (\ref{anticom}):
 \begin{equation}
 Q^{\dagger}_a Q^{\dagger}_b Q^{\dagger}_c + Q^{\dagger}_b Q^{\dagger}_c Q^{\dagger}_a + Q^{\dagger}_c Q^{\dagger}_a Q^{\dagger}_b
 = {\bar{\eta}}_{ a b c } \, {\mbox{l\hspace{-0.55em}1}}, \; \; \; a, b,... = 1,2,3.
\label{anticomdot}
\end{equation}
with ${\bar{\eta}}_{abc}$ complex conjugate of $\eta_{abc}$.

It is obvious that any similarity transformation of the generators $Q_a$ will keep the ternary anti-commutator (\ref{defeta})
invariant. As a matter of fact, if we define ${\tilde{Q}}_b = P^{-1} Q_b P$, with $P$ a non-singular $3 \times 3$ matrix,
the new set of generators will satisfy the same ternary relations, because 
$${\tilde{Q}}_a {\tilde{Q}}_b {\tilde{Q}}_c = P^{-1} Q_a P P^{-1} Q_b P P^{-1} Q_c P = P^{-1} (Q_a Q_b Q_c) P,$$
and on the right-hand side we have the unit matrix which commutes with all other matrices, 
so that $P^{-1} \; {\mbox{l\hspace{-0.55em}1}} \; P  ={\mbox{l\hspace{-0.55em}1}}$.

The six matrices $Q_a$ and $Q^{\dagger}_b, \; \; a,b,...=1,2,3$ are traceless, and one can define 
six traceless {\it hermitian} matrices.
forming the following linear combinations: $\frac{1}{2} \; ( Q_a + Q^{\dagger}_a)$ and $\frac{1}{2i} \; ( Q_a - Q^{\dagger}_a ) $.

This is not enough to produce the complete basis for $3 \time 3$ traceless hermitian matrices, which should be of dimension $8$.
Two linearly independent traceless {\it diagonal} matrices must be added; we choose the following:
\begin{equation}
B = \begin{pmatrix} 1 & 0 & 0 \cr 0 & j & 0 \cr 0 & 0 & j^2 \end{pmatrix} \; \; \; \; {\rm and} \; \; \; 
B^{\dagger} = B^2 = \begin{pmatrix} 1 & 0 & 0 \cr 0 & j^2 & 0 \cr 0 & 0 & j \end{pmatrix}.
\label{BBmatrices}
\end{equation} 
One can also easily check that 
\begin{equation}
B \, B^{\dagger} =  B^{\dagger} \, B = {\mbox{l\hspace{-0.55em}1}}, \; \; \; \left( B^{\dagger} \right)^2 = B, \; \; \; \;  B^3 = {\bf 1} \; \; \; 
{\rm and} \left( B^{\dagger} \right)^3 = {\mbox{l\hspace{-0.55em}1}}.
\label{BBmat2}
 \end{equation}
The set of eight traceless matrices $(Q_a, Q^{\dagger}_b, B$ and $B^{\dagger})$ forms an associative algebra over the 
ring of real numbers tensorized with the groups $Z_3$ generated by the complex third root of unity $j=e^{\frac{2 \pi i}{3}}$.
These matrices can serve as a basis (although unusual) of the $SU(3)$ Lie algebra, see e.g. \cite{VKac1994}. The matrices
of this type were used recently in the description of Yangians by Yu and Ge (\cite{LiWei}).

The energy operator, proportional to the $12 \times 12$ unit matrix, can be written in a similar manner
as a product of three unit matrices, 
${\mbox{l\hspace{-0.55em}1}}_3 \otimes {\mbox{l\hspace{-0.55em}1}}_2 \otimes {\mbox{l\hspace{-0.55em}1}}_2$ 

In the basis in which the functions are aligned in a column by colors, first $\varphi_{+}, \chi_{+}, \psi_{+}$
followed by $\varphi_{-}, \chi_{-}, \psi_{-}$, the matrix operators take on another form, namely
 \begin{equation}
M = m \; \sigma_3 \otimes B \otimes {\bf 1}_2, \; \; \; \; 
P =  \sigma_1 \otimes Q_3 \otimes {\boldsymbol{\sigma}} \cdot {\bf p} 
\label{MPtensor2}
\end{equation}
Keeping only the mass operator on the right-hand side, we get:
\begin{equation}
 E \; \left[{\bf 1}_2 \otimes {\bf 1}_3 \otimes {\bf 1}_2 \right] \; \Psi -
c \; \left[ \sigma_1 \otimes Q_3 \otimes {\boldsymbol{\sigma}} \cdot {\bf p} \right] \; \Psi
= mc^2 \; \left[ \sigma_3 \otimes B \otimes {\bf 1}_2 \right] \; \Psi
\label{Bigpsi}
\end{equation}
By multiplying on the left by the matrix 
$$ \sigma_3 \otimes B^{\dagger} \otimes {\bf 1}_2$$
we arrive at the following form of ternary generalization of Dirac's equation:
$$ E \; \left[ \sigma_3 \otimes B^{\dagger} \otimes {\bf 1}_2 \right] -
\left[ c \; i \sigma_2 \otimes j^2 \,Q_2 \otimes {\boldsymbol{\sigma}} \cdot {\bf p} \right] \;
\Psi= mc^2 \; {\bf 1}_2 \otimes {\bf 1_3} \otimes {\bf 1}_2 \; \Psi$$
where we used the fact that under matrix multiplication, $\sigma_3 \sigma^3 = {\mbox{l\hspace{-0.55em}1}}_2$,
$B^{\dagger} B = {\mbox{l\hspace{-0.55em}1}}_3$ and $B^{\dagger} Q_3 = j^2 \, Q_2$.

One can check by direct computation that the sixth power of this operator gives the same result as before,
\begin{equation}
\left[ E \;  \sigma_3 \otimes B^{\dagger} \otimes {\mbox{l\hspace{-0.55em}1}}_2
- i \sigma_2 \otimes j^2 \, Q_2 \otimes c \, {\boldsymbol{\sigma}}\cdot {\bf p} \right]^6 =
\left[ E^6 - c^6 {\bf p}^6 \right] \; {\mbox{l\hspace{-0.55em}1}}_{12} = m^6 c^{12} \;  {\mbox{l\hspace{-0.55em}1}}_{12}
\label{sixpower1}
\end{equation}
The ternary Dirac equation can be written in a concise manner using the Minkowskian indices 
and the usual pseudo-scalar product of two four-vectors as follows:
\begin{equation}
\Gamma^{\mu} p_{\mu} = m c^2 \; {\mbox{l\hspace{-0.55em}1}}_{12}
\label{Gammasecond}
\end{equation}
with $12 \times 12$ matrices $\Gamma^{\mu}, \mu = 0, 1, 2, 3$ defined as follows:
$$\Gamma^0 = \sigma_3 \otimes B^{\dagger} \otimes {\mbox{l\hspace{-0.55em}1}}_2, \; \; \; \; 
\hskip 0.6cm 
\; \; \; \; \Gamma^{k} = - i \sigma_2 \otimes j^2 \, Q_2 \otimes  {\sigma}^k $$

\section{The $Z_3$ Lorentz symmetry}

Let us rewrite the matrix operator generating our system when it acts on the column vector
containing twelve components of three ``color" fields,
$$E \; {\mbox{l\hspace{-0.55em}1}}_2 \otimes {\mbox{l\hspace{-0.55em}1}}_3 \otimes {\mbox{l\hspace{-0.55em}1}}_2 
= m c^2 \; \sigma_3 \otimes B \otimes {\mbox{l\hspace{-0.55em}1}}_2
 + \sigma_1 \otimes Q_3 \otimes c \, {\boldsymbol{\sigma}}\cdot {\bf p}$$
in a slightly different way, with energy and momentum operators on the left hand side, and the mass operator
on the right hand side:
\begin{equation}
E \; {\mbox{l\hspace{-0.55em}1}}_2 \otimes {\mbox{l\hspace{-0.55em}1}}_3 \otimes {\mbox{l\hspace{-0.55em}1}}_2 
- \sigma_1 \otimes Q_3 \otimes {\boldsymbol{\sigma}}\cdot {\bf p}
= m c^2 \; \sigma_3 \otimes B \otimes {\mbox{l\hspace{-0.55em}1}}_2
%+ \sigma_1 \otimes Q_3 \otimes c \, {\boldsymbol{\sigma}}\cdot {\bf p}$$
\label{EPtogether}
\end{equation}
Following a similar procedure applied to the Dirac equation, let us transform this equation
so that the mass operator becomes proportional to the unit matrix. Let us multiply this equation
 from the left by the matrix $\sigma_3 \otimes B^{\dagger} \otimes {\mbox{l\hspace{-0.55em}1}}_2.$ 

Now we get the following equation which enable us to interpret the energy and the momentum as the components of 
a Minkowskian four-vector $c \; p^{\mu} = [E, \; c {\bf p}] $:
\begin{equation}
E \;  \sigma_3 \otimes B^{\dagger} \otimes {\mbox{l\hspace{-0.55em}1}}_2
- i \sigma_2 \otimes j^2 \, Q_2 \otimes c \, {\boldsymbol{\sigma}}\cdot {\bf p} = 
m c^2 \;  {\mbox{l\hspace{-0.55em}1}}_2 \otimes {\mbox{l\hspace{-0.55em}1}}_3 \otimes {\mbox{l\hspace{-0.55em}1}}_2,
\label{Gammafirst}
\end{equation}
where we used the fact that under matrix multiplication, $\sigma_3 \sigma^3 = {\mbox{l\hspace{-0.55em}1}}_2$,
$B^{\dagger} B = {\mbox{l\hspace{-0.55em}1}}_3$ and $B^{\dagger} Q_3 = j^2 \, Q_2$.

%\begin{frame}{Diagonalization at the $6$-th power}

One can check by direct computation that the sixth power of this operator gives the same result as before,
$$\left[ E \;  \sigma_3 \otimes B^{\dagger} \otimes {\mbox{l\hspace{-0.55em}1}}_2
- i \sigma_2 \otimes j^2 \, Q_2 \otimes c \, {\boldsymbol{\sigma}}\cdot {\bf p} \right]^6 =$$
\begin{equation}
= \left[ E^6 - c^6 {\bf p}^6 \right] \; {\mbox{l\hspace{-0.55em}1}}_{12} = m^6 c^{12} \; {\mbox{l\hspace{-0.55em}1}}_{12}
\label{sixpower2}
\end{equation}
It is also worthwhile to note that not only taking the sixth power of our operator yields the simple algebraic relation 
(\ref{sixpower2}), but the similar relation exists between the determinants: 
\begin{equation}
{\rm det} \left( E \;  \sigma_3 \otimes B^{\dagger} \otimes {\mbox{l\hspace{-0.55em}1}}_2
- i \sigma_2 \otimes j^2 \, Q_2 \otimes c \, {\boldsymbol{\sigma}}\cdot {\bf p} =  \right) = \left( E^6 - c^6 \mid {\bf p} \mid^6 \right)^2
= {\rm det} \left( m c^2 \;  {\mbox{l\hspace{-0.55em}1}}_2 \otimes {\mbox{l\hspace{-0.55em}1}}_3 \otimes {\mbox{l\hspace{-0.55em}1}}_2, \right)
= m^{12} c^{24}.
\label{Dets12}
\end{equation}
The eigenvalues of the generalized Dirac operator have all the same absolute value $R = \mid (E^6 - c^6 \mid {\bf p} \mid^6)^{\frac{1}{6}} \mid$,
and are given by:
\begin{equation}
R, \; -R, \; j R, \; -j R, \; j^2 R, \; -j^2 R.
\label{Eigen6}
\end{equation}
They are double degenerate, i.e. although the characteristic equation is of twelfth order, it has only six distinct eigenvalues. 
This result will be important for the subsequent discussion of the generalized Lorentz invariance.

Our equation can be written in a concise manner using the Minkowskian indices 
and the usual pseudo-scalar product of two four-vectors as follows:
\begin{equation}
\Gamma^{\mu} p_{\mu} = m c \; {\mbox{l\hspace{-0.55em}1}}_{12}, \; \; \; {\rm with} \; \; p^0 = \frac{E}{c}, \; \; p^k = m c \frac{d x^k}{ds}.
\label{GammasecondB}
\end{equation}
with $12 \times 12$ matrices $\Gamma^{\mu}, \mu = 0, 1, 2, 3$  defined as follows:
\begin{equation}
\Gamma^0 = \sigma_3 \otimes B^{\dagger} \otimes {\mbox{l\hspace{-0.55em}1}}_2, \; \; \; \; \; \; 
\Gamma^{k} = - i \sigma_2 \otimes j^2 \, Q_2 \otimes  {\sigma}^k
\label{Gammamu}
\end{equation}
Unfortunately, the four $12 \times 12$ matrices do not satisfy usual anti-commutation relations 
similar to those of the  $4 \times 4$ Dirac matrices $\gamma^{\mu}$, i.e.
$\gamma^{\mu} \gamma^{\nu} + \gamma^{\nu} \gamma^{\mu} = 2 \; g^{\mu \nu} \; {\bf 1}_4.$ 

Although the four $12 \times 12$ matrices do not satisfy usual anti-commutation relations 
similar to those of the $4 \times 4$ Dirac matrices $\gamma^{\mu}$, i.e.
$\gamma^{\mu} \gamma^{\nu} + \gamma^{\nu} \gamma^{\mu} = 2 \; g^{\mu \nu} \; {\bf 1}_4.$ 
nevertheless, the system of equations satisfied by the 12-dimensional wave function $\Psi$,
\begin{equation}
-i \hbar \; \Gamma^{\mu} \, \partial_{\mu} \, \Psi = m c \Psi
\label{TernDirac}
\end{equation}
is a hyperbolic one, and has the same light cone as the Klein-Gordon equation. To corroborate this statement, let us
first consider the massless case,
\begin{equation}
 -i \hbar \; \Gamma^{\mu} \, \partial_{\mu} \, \Psi = 0.
\label{TernDirac0}
\end{equation}
Assuming the general solution of the form $e^{k_{\mu} x^{\mu}}$, we can replace the derivations by
the components of the wave 4-vector $k^{\mu}$, and take the sixth power of the matrix $\Gamma^{\mu} k_{\mu}$. 
The resulting dispersion relation was shown to be
$$
k_0^6 - \mid {\bf k} \mid^6 = \left( k_0^2 - \mid {\bf k} \mid^2 \right)  \left( k_0^2 - j \; \mid {\bf k} \mid^2 \right)
 \left( k_0^2 - j^2 \; \mid {\bf k} \mid^2 \right)=$$
$$ =\left( k_0^2 - \mid {\bf k} \mid^2 \right) 
\; \left( k_0^4 + k_0^2  \mid {\bf k} \mid^2 + \mid {\bf k} \mid^4 \right) = 0.$$
The first factor defines the usual light cone, while the factor of degree four is strictly positive
(besides the origin $0$). 
The system has only one characteristic surface which is the same for all massless fields.
Each of the three factors remains invariant under a different representation of the $SL (2, {\bf C})$ group.

Let us introduce the following three matrices representing the same four-vector $k^{\mu}$:
\begin{equation}
K_3 = \begin{pmatrix} k_0 & k_x \cr k_x & k_0 \end{pmatrix}, \; \; K_1 = \begin{pmatrix} k_0 & j  k_x \cr j k_x & k_0 \end{pmatrix},
\; \; K_2 = \begin{pmatrix} k_0 & j^2  k_x \cr j^2 k_x & k_0 \end{pmatrix},
\label{threekmat}
\end{equation}  
whose determinants are, respectively,
\begin{equation}
{\rm det} K_1 = k_0^2 - j^2 k_x^2, \; \; \; {\rm det} K_2 = k_0^2 - j k_x^2, \; \; {\rm det} K_3 = k_0^2 -  k_x^2.
\label{threedets}
\end{equation} 

Note that only the third matrix $K_3$ is hermitian,
 and corresponds to a {\it real} space-time vector $k^{\mu}$,
while neither of the remaining two matrices $K_1$ and $K_2$ is hermitian; 
however, one is the hermitian conjugate of another.

In what follows, we shall replace the absolute value of the wave vector $\mid {\bf k} \mid$ 
 by a single spatial component, say $k_x$,
because for any given $4$-vector $k^{\mu} = [k_0, {\bf k} ]$  we can choose the coordinate system in such a way that 
its $x$-axis should be aligned along the vector ${\bf k}$.
Then it is easy to check that one has:

$$
\begin{pmatrix} \cosh u & \sinh u \cr \sinh u & \cosh u \end{pmatrix} \begin{pmatrix} k_0 \cr k_x \end{pmatrix} = 
\begin{pmatrix} {k'}_0 \cr {k'}_x \end{pmatrix} $$
$$
\begin{pmatrix} \cosh u & j^2 \sinh u \cr j \sinh u & \cosh u \end{pmatrix} \begin{pmatrix} k_0 \cr j \; k_x \end{pmatrix} = 
\begin{pmatrix} {k'}_0 \cr j \; {k'}_x \end{pmatrix} $$ 
\begin{equation}
\begin{pmatrix} \cosh u & j \sinh u \cr j^2 \sinh u & \cosh u \end{pmatrix} \begin{pmatrix} k_0 \cr j^2 k_x \end{pmatrix} = 
\begin{pmatrix} {k'}_0 \cr j^2 {k'}_x \end{pmatrix} 
\end{equation}
The transformed vectors are given by the following expressions:

$$ \hskip 0.5cm i) \; \;  k^{'}_0 = k_0  \cosh u + k_x \; \sinh u, \; \; \; k^{'}_x =   k_0  \sinh u +  k_x \; \cosh u $$

$$\hskip 0.3cm  ii) \; \; k^{'}_0 = k_0  \cosh u + j^2 \; k_x \; \sinh u, \; \; \; k^{'}_x =  j \; k_0  \sinh u +  k_x \; \cosh u $$

$$iii) \; \; k^{'}_0 = k_0  \cosh u + j \; k_x \; \sinh u, \; \; \; k_x^{'} =  j^2 \; k_0  \sinh u +  k_x \; \cosh u $$
 Let us now introduce a $6 \times 6$ matrix composed out of the above three $2 \times 2$ matrices: 
 \begin{equation}
\begin{pmatrix} 0 & k_0 \; {\mbox{l\hspace{-0.55em}1}}_{2} + {\bf k} \cdot {\boldsymbol{\sigma}} & 0 \cr 
 0 & 0 &  k_0 \; {\mbox{l\hspace{-0.55em}1}}_{2} + j \, {\bf k} \cdot {\boldsymbol{\sigma}}  \cr
k_0 \; {\mbox{l\hspace{-0.55em}1}}_{2}+ j^2 \, {\bf k} \cdot {\boldsymbol{\sigma}}& 0 & 0 \end{pmatrix}
\label{threekays}
\end{equation}
or, more explicitly,
\begin{equation}
{\Large K} = \begin{pmatrix}  0 & 0 & k_0 & k_x & 0 & 0 \cr 0 & 0 & k_x & k_0 & 0 & 0 \cr
0 & 0 & 0 & 0 & k_0 & j k_x  \cr 0 & 0 & 0 & 0 &  j k_x & k_0  \cr  k_0 & j^2 k_x & 0 & 0 & 0 & 0 \cr
j^2 k_x & k_0 & 0 & 0 & 0 & 0 \end{pmatrix} 
\label{bigmatrix}
\end{equation} 
It is easy to check that 
$$
{\rm det \Large K} = \left( {\rm det} K_1 \right) \cdot \left( {\rm det} K_2 \right) \cdot \left( {\rm det} K_3 \right) $$
\begin{equation}
= (k_0^2 - k_x^2)(k_0^2 - j^2 k_x^2) ( k_0^2 - j k_x^2) = k_0^6 - k_x^6.
\label{bigdet}
\end{equation}
It is also remarkable that the determinant remains the same in the basis in which 
the ternary Dirac operator was proposed, namely when we consider the matrix 
\begin{equation}
{\Large K} = \begin{pmatrix}  k_0 & 0 & 0 & k_x & 0 & 0 \cr 0 & k_0 & k_x & 0 & 0 & 0 \cr
0 & 0 & k_0 & 0 & 0 & j k_x  \cr 0 & 0 & 0 & k_0 &  j k_x & 0  \cr  0 & j^2 k_x & 0 & 0 & k_0 & 0 \cr
j^2 k_x & 0 & 0 & 0 & 0 & k_0 \end{pmatrix} 
\label{bigmatrix2}
\end{equation} 
 Let us show now that the spinorial representation of Lorentz boosts can be applied 
 to each of the three matrices $K_1, K_2$ and $K_3$
separately, keeping their determinants unchanged.
As a matter of fact, besides the well-known formula:
\begin{equation}
\begin{pmatrix} \cosh \frac{u}{2} & \sinh \frac{u}{2} \cr \sinh \frac{u}{2} & \cosh \frac{u}{2} \end{pmatrix}
 \hskip 0.2cm \begin{pmatrix} k_0 & k_x \cr k_x & k_0 \end{pmatrix} \hskip 0.2cm 
\begin{pmatrix} \cosh \frac{u}{2} & \sinh \frac{u}{2} \cr \sinh \frac{u}{2} & \cosh \frac{u}{2} \end{pmatrix} = 
 \begin{pmatrix} {k'}_0 & {k'}_x \cr {k'}_x & {k'}_0 \end{pmatrix},
\label{spinlorentz1} 
\end{equation}
with
\begin{equation}
{k'}_0 = k_0 \; \cosh u + k_x \; \sinh u, \; \; \; \; {k'}_x = k_0 \; \sinh u + k_x \; \cosh u.
\label{kprimone}
\end{equation}
which becomes apparent when we remind that 
$$\cosh^2 \frac{u}{2} + \sinh^2 \frac{u}{2} = \cosh u \;\; {\rm and} \; \;  2 \sinh \frac{u}{2} \; \cosh \frac{u}{2} = \sinh u,$$ 
keeping unchanged the Minkowskian scalar product: ${k'}^2_0 - {k'}_x^2 = k_0^2 - k_x^2$, 
we have also two transformations of the same kind which keep invariant the ``complexified" Minkowskian squares appearing
as factors in the sixth-orer expression $k_0^6 - k_x^6$, namely 
$$ k_0^2 - j \; k_x^2 \; \; \; \; {\rm and} \; \; \;  k_0^2 - j^2 \; k_x^2. $$
The above expressions can be identified as the determinants of the following $2 \times 2$ matrices: 
\begin{equation}
k_0^2 - j \; k_x^2 = {\rm det} \begin{pmatrix} k_0 & j^2 k_x^2 \cr j^2 \; k_x & k_0 \end{pmatrix}, \; \; \; \;  
k_0^2 - j^2 \; k_x^2 = {\rm det} \begin{pmatrix} k_0 & j k_x^2 \cr j \; k_x & k_0 \end{pmatrix}.
\label{twokinv}
\end{equation}  
It is easy to check that we have: 
\begin{equation}
\begin{pmatrix} \cosh \frac{u}{2} & \sinh \frac{u}{2} \cr \sinh \frac{u}{2} & \cosh \frac{u}{2} \end{pmatrix}
  \begin{pmatrix} k_0 & j k_x \cr j k_x & k_0 \end{pmatrix} 
\begin{pmatrix} \cosh \frac{u}{2} & \sinh \frac{u}{2} \cr \sinh \frac{u}{2} & \cosh \frac{u}{2} \end{pmatrix} = 
 \begin{pmatrix} {k'}_0 & j {k'}_x \cr j {k'}_x & {k'}_0 \end{pmatrix},
\label{spinlorentz2} 
\end{equation}
with  ${k'}_0 = k_0 \; \cosh u + j \; {k'}_x \; \sinh u $, so that ${k'}_0^2 - j {k'}_x^2 = k_0^2 - j k_x^2$,  
 as well as 
\begin{equation}
\begin{pmatrix} \cosh \frac{u}{2} & \sinh \frac{u}{2} \cr \sinh \frac{u}{2} & \cosh \frac{u}{2} \end{pmatrix}
\begin{pmatrix} k_0 &j^2 \; k_x \cr j^2 \; k_x & k_0 \end{pmatrix} 
\begin{pmatrix} \cosh \frac{u}{2} & \sinh \frac{u}{2} \cr \sinh \frac{u}{2} & \cosh \frac{u}{2} \end{pmatrix} = 
 \begin{pmatrix} {k'}_0 & j^2 {k'}_x \cr j^2 {k'}_x & {k'}_0 \end{pmatrix},
\label{spinlorentz3} 
\end{equation}
The bottom line is the following: the $12 \times 12$ matrix formed by the tensor product of $\sigma_3$ with the 
$6 \times 6$ matrix $K$ defined above, has the same determinant and the same eigenvalues (\ref{Dets12}, \ref{Eigen6})
as the generalized Dirac operator \ref{EPtogether}, if we replace $k_0$ by $E$ and ${\bf k}$ by $c {\bf p}$.
We have shown that the determinant of the matrix $\sigma_3 \otimes K$ (equal to $(k_0^6 - k_x^6)^2$ remains
invariant under the generalized Lorentz transformation composed of three representations, the usual unitary one
and two complex ones. Therefore there exists a similarity between the two matrices, which preserves the invariance
under the generalized Lorentz group intertwined with $Z_3$. 

\section{Interaction with gauge fields}

The matrix representation of the system (\ref{EPtogether}) is by no means unique. In the form which most closely
resembles the classical Dirac equation, we chose the following representation for our ternary Dirac operator (designed be 
${\cal{D}}$ for convenience):
\begin{equation}
{\cal{D}} = E \; \sigma_3 \otimes B^{\dagger} \otimes {\mbox{l\hspace{-0.55em}1}}_2 
- (i \sigma_2 \otimes j Q_2 \otimes c \; {\boldsymbol{\sigma}} \cdot {\bf p} =
mc^2 \; {\mbox{l\hspace{-0.55em}1}}_2 \otimes {\mbox{l\hspace{-0.55em}1}}_3 \otimes {\mbox{l\hspace{-0.55em}1}}_2
\label{Dirac3last}
\end{equation}
Obviously, the essential sixth order diagonalized system resulting from the sixth iteration of this operator, as
well as its characteristic equation and eigenvalues remain unchanged under an arbitrary similarity transformation,
${\cal{D}} \rightarrow P^{-1} {\cal{D}} P$. Taking into account the particular tensorial structure of ternary Dirac operator,
the matrices $P$ should display similar structure in order to keep the three factors separated. This reduces the allowed
similarity matrices to the following family:
$$P = R \otimes S \otimes U, \; \; $$
with $R$ being a $2 \times 2$ matrix, $S$ denoting a $3 \times 3$ matrix, and $U$ proportional to the $2 \times 2$
unit matrix in order not to change the scalar product ${\boldsymbol{\sigma}} \cdot {\bf p}$ in the last tensorial factor
in ${\cal{D}}$. 

The minimal coupling between the Dirac particles (electrons and positrons) with the electromagnetic field is obtained by
inserting the four-potential $A_{\mu}$  into the Dirac equation:
\begin{equation}
\gamma^{\mu} (p_{\mu} - e \; A_{\mu} ) \; \psi = m \; \psi.
\label{DiracwithA}
\end{equation}
Ternary generalization of Dirac's equation, when expressed  with explicit Minkowskian
indices, offers a similar possibility of introducing gauge fields. The particular structure of $12 \times 12$ matrices 
$\Gamma_{\mu}$  makes possible the accomodation of three types of gauge fields, corresponding to three factors 
from which the tensor product results.

The overall gauge field can be decomposed into a sum of three contributions:
the $SU(3)$ gauge field $\lambda_a B^a_{\mu}$, with $\lambda_a, \; \; \; a = 1,2,..8$
 denoting the eight $3 \times 3$ traceless Gell-Mann matrices, the $SU(2)$ gauge field 
$\sigma_k \, A^k_{\mu}, \; \; \; k=1,2,3$ and the electric field potential $A_{\mu}$. 
We propose to insert each of these gauge potentials into a common $12 \times 12$ matrix as follows:
The strong interaction gauge potential is aligned on the $SU(3)$ matrix basis:
$$B_{\mu} = {\mbox{l\hspace{-0.55em}1}}_2 \otimes {\lambda}_a B^a_{\mu} \otimes {\mbox{l\hspace{-0.55em}1}}_2, \; \; , a, b =1,2,...8. $$
appearing as the second factor in the tensor product;

The $SU(2)$ weak interaction potential $A^i_{\mu}$ aligned along the three $\sigma$-matrices of the first tensorial factor
$$\sigma_k \, A^k_{\mu} \otimes {\mbox{l\hspace{-0.55em}1}}_3, \; \; \;  i, k, ..= 1,2,3.$$
and the electromagnetic potential $A^{em}_{\mu}$ aligned along the unit  $2 \times 2$ matrix appearing as the
third factor in the tensor product.
$$\otimes A_{\mu} \, {\mbox{l\hspace{-0.55em}1}}_2 $$
so that the overall expression for the gauge potential becomes:

\begin{equation}
{\cal{A}}_{\mu} = {\mbox{l\hspace{-0.55em}1}}_2 \otimes {\lambda}_a B^a_{\mu} \otimes {\mbox{l\hspace{-0.55em}1}}_2 +
\sigma_k \, A^k_{\mu} \otimes {\mbox{l\hspace{-0.55em}1}}_3 \otimes {\mbox{l\hspace{-0.55em}1}}_2 +
 {\mbox{l\hspace{-0.55em}1}}_2 \otimes {\mbox{l\hspace{-0.55em}1}}_3 \otimes A^{em}_{\mu} \, {\mbox{l\hspace{-0.55em}1}}_2 
\end{equation}
%\label{Bigpotential}
%\end{equation}
The proposed ternary generalization of Dirac's equation including color degrees of freedom contains naturally not only
the $SU(3)$-invariant strong interactions, but leads automatically to another type of gauge fields to which quarks are
also sensitive: these are the gauge fields generated by the $SU(2)$ and $U(1)$ symmetries incorporated in the system.

There is an extra bonus here: namely, one can look at the same system (\ref{EPtogether}) in the limit when the color interaction
is switched off. This amounts to replacing the $3 \times 3$ matrices $B$ and $Q_3$ by unit matrices ${\mbox{l\hspace{-0.55em}1}}_3.$
The resulting system is equivalent with a cartesian product of three identical Dirac equations:
\begin{equation}
E \; {\mbox{l\hspace{-0.55em}1}}_2 \otimes {\mbox{l\hspace{-0.55em}1}}_3 \otimes {\mbox{l\hspace{-0.55em}1}}_2 
- \sigma_1 \otimes {\mbox{l\hspace{-0.55em}1}}_3 \otimes {\boldsymbol{\sigma}}\cdot {\bf p}
= m c^2 \; \sigma_3 \otimes {\mbox{l\hspace{-0.55em}1}}_3 \otimes {\mbox{l\hspace{-0.55em}1}}_2
%+ \sigma_1 \otimes Q_3 \otimes c \, {\boldsymbol{\sigma}}\cdot {\bf p}$$
\label{Nocolors}
\end{equation}
Without any symmetry breaking, this set of equations describes three identical fermions sensitive exclusively to the
 $SU(2) \times U(1)$ gauge fields, i.e. the electroweak interaction, like the elementary particles known as {\it leptons} - in this setting they
appear as natural {\it colorless} companions of quarks. This sheds new light on the fact that their number is equal,
and even if other families of quarks had to be introduced (which we did not consider here), described by a similar 
ternary Dirac system, they would also give rise to another set of three leptons. And this is what the experimental data
confirmed since the discovery of the families with other ``flavors". The gauge fields are obviously common to all families.

In principle, we should have started with zero masses for all particles, quarks and leptons alike, and let the Higgs-Kibble mechanism 
generate non-zero masses. The Higgs field necessary for this to happen can be introduced like in the model of matrix algebras
in the context of {\it non-commutative geometry}, (see \cite{MDVRKJM1}, \cite{MDVRKJM2}, \cite{Shadow}; see also \cite{Kerner1983}).

\section{Solutions}

%Applying the quantum correspondence principle, the equation relating mass, energy and momentum:
%\begin{equation}
%E^6 \, u(t, {\bf r}) = m^6 c^{12} \, u(t, {\bf r}) + c^6 \, {\bf p}^6 \, u(t, {\bf r})
%\label{u-equation}
%\end{equation}
%is transformed into a linear differential equation of the sixth order. Indeed, according to
%\begin{equation}
%E \rightarrow - i \, \hbar \frac{\partial}{\partial t}, \; \; \; \; 
%{\bf p} \rightarrow - i \, \hbar \, {\boldsymbol{\nabla}},
%\label{correspondence}
%\end{equation}
%
%For given $\omega$ and ${\bf k}$ (corresponding to the energy $E = \hbar \, \omega$ and momentum
%${\bf p} = \hbar \, {\bf k}$), the general solution of the equation (\ref{

The system  of twelve linear equations supposed to describe the dynamics of three 
intertwined fields was shown to be represented by a single matrix operator acting on a $12$-component vector: symbolically
$E \Psi = (c^2 M + c P) \Psi.$
By consecutive application of this matrix operator we are able separate the variables 
and find the common equation of sixth order that is satisfied by each of the components:
\begin{equation}
E^6 \Psi = m^6 c^{12} \Psi + c^6 \, {\bf p}^6 \Psi.
%E^6 \, u(t, {\bf r}) = m^6 c^{12} \, u(t, {\bf r}) + c^6 \, {\bf p}^6 \, u(t, {\bf r})
\label{u-equation}
\end{equation}
Applying the quantum correspondence principle, the above equation relating mass, energy and momentum (\ref{u-equation})
%\begin{equation}
%E^6 \, u(t, {\bf r}) = m^6 c^{12} \, u(t, {\bf r}) + c^6 \, {\bf p}^6 \, u(t, {\bf r})
%\label{u-equation}
%\end{equation}
is transformed into a linear differential equation of the sixth order. Indeed, according to 
\begin{equation}
E \rightarrow - i \, \hbar \frac{\partial}{\partial t}, \; \; \; \; 
{\bf p} \rightarrow - i \, \hbar \, {\boldsymbol{\nabla}},
\label{correspondence}
\end{equation}
we get the following sixth-order partial differential equation to be satisfied 
by all the components of the wave function $\Psi$. 
\begin{equation}
- {\hbar}^6 \frac{\partial^6}{\partial t^6} \psi - m^6 c^{12} \psi = - {\hbar}^6 {\Delta}^3 \psi.
\label{sixthorder}
\end{equation}
Identifying quantum operators of energy and momentum,
$$ - i {\hbar} \frac{\partial}{\partial t} \rightarrow E, \; \; \; - i {\hbar} {\bf \nabla} \rightarrow {\bf p},$$
Let us write the algebraic expression relating mass, energy and momentum (\ref{u-equation}) simply as follows:
\begin{equation}
E^6 - m^6 c^{12} = \mid {\bf p} \mid^6 c^6.
\label{Ep_relation}
\end{equation}
This equation can be factorized showing how it was obtained by subsequent action of 
the operators of the system of six equations:
{\small
$$
E^6 - m^6 c^{12} = (E^3 - m^3 c^6)(E^3 + m^3 c^6) =$$
$$
 (E - mc^2)(jE - mc^2)(j^2 E - mc^2)(E + mc^2)(jE + mc^2)(j^2 E + mc^2) = \mid {\bf p} \mid^6 c^6.
$$}
This sixth-order equation can be solved by separation of variables; the time-dependent
and the space-dependent factors have the same structure:
$$A_1 \,e^{\omega\,t} + A_2 \,e^{j \,\omega\,t} + A_3 e^{j^2 \,\omega\,t},\,
\ \ \ \ B_1\,e^{{\bf k.r}} + B_2\,e^{j\,{\bf k.r}} + B_3\,e^{j^2\,{\bf k.r}}$$
with $\omega$ and ${\bf k}$ satisfying the following dispersion relation:
\begin{equation}
\frac{\omega^6}{c^6} + \frac{m^6 c^6}{{\hbar}^6} = \mid {\bf k} \mid^6,
\label{dispersion6}
\end{equation}
where we have identified  $E = {\hbar \omega}$  and ${\bf p} = {\hbar} {\bf k}$. 

Up to this point we follow exactly the way in which the Klein-Gordon equation is deduced from the
Dirac equation as the common condition to be satisfied by all the components of the Dirac spinor:
\begin{equation}
E^2 \psi = m^2 c^4 \psi + c^2 {\bf p}^2 \psi \; \; \rightarrow \; \; 
-\hbar^2 \frac{\partial^2 \psi}{\partial t^2} = m^2 c^4 \psi - \hbar^2 \triangle \psi.
\label{KGequation}
\end{equation}
The solutions  are saught in the plane wave form $\psi \sim e^{i(\omega t + {\bf k}\cdot{\bf r})}$.
 Due to the purely imaginary exponential, after such a substitution the Klein-Gordon equation reduces to the well known 
algebraic condition 
\begin{equation}
\hbar^2 \omega^2 = m^2 c^4 + \hbar^2 {\bf k}^2,
\label{omegahbar}
\end{equation}
which coincides with the previously established relation between the energy, momentum and mass due to the
correspondence $E = \hbar \omega$ and ${\bf p} = \hbar {\bf k}$ introduced by de Broglie. 

The sixth-order dispersion relation  is invariant under
the action of $Z_2 \times Z_3 = Z_6$ symmetry, because to any solution with given real
$\omega$ and ${\bf k}$ one can add solutions with $\omega$ replaced by $j \omega$ or $j^2 \omega$, $j {\bf k}$ 
or $j^2 {\bf k}$, as well as  $- \omega$; there is no need to introduce also $- {\bf k}$ instead of ${\bf k}$
because the vector ${\bf k}$ can take on all possible directions covering the unit sphere.

The nine complex solutions with positive frequency $\omega$ as well as with $j \; \omega$
and $j^2 \; \omega$ obtained by the action of the $Z_3$-group can be displayed in a compact manner
in form of a $3 \times 3$ matrix. The inclusion of the essential $Z_2$-symmetry ensuring the 
existence of {\it anti-particles} leads to the nine similar solutions with negative $\omega$. 
The two matrices are displayed below:
$$
\begin{pmatrix}  e^{\omega\,t + {\bf k \cdot r}} & e^{\omega\,t + j {\bf k \cdot r}}
& e^{\omega\,t + j^2 {\bf k \cdot r}} \cr 
e^{j \omega\,t + {\bf k \cdot r }} & 
e^{j \omega\,t + j {\bf k \cdot r}} &
e^{j \omega\,t + j^2 {\bf k \cdot r}} \cr
e^{j^2 \omega\,t + {\bf k \cdot r}}& e^{j^2 \omega\,t + {\bf k \cdot r}}  & 
e^{j^2 \omega\,t + j^2 {\bf k \cdot r}} 
\end{pmatrix}, \; \; \; \; \; 
\begin{pmatrix} e^{- \omega\,t - {\bf k \cdot r}} & e^{-\omega\,t - j {\bf k \cdot r}}
& e^{-\omega\,t - j^2 {\bf k \cdot r}} \cr 
e^{-j \omega\,t-{\bf k \cdot r }} & 
e^{-j \omega\,t - j {\bf k \cdot r}} &
e^{-j \omega\,t - j^2 {\bf k \cdot r}} \cr
e^{-j^2 \omega\,t -{\bf k \cdot r}}& e^{-j^2 \omega\,t - {\bf k \cdot r}}  & 
e^{-j^2 \omega\,t - j^2 {\bf k \cdot r}} 
\end{pmatrix} $$
and their nine {\it real} linear combinations can be represented in the following  $3 \times 3$
 matrix of functions as follows:
{\small $$
\begin{pmatrix}   e^{\omega\,t + {\bf k \cdot r}} &   e^{\omega\,t - \frac{{\bf k \cdot r}}{2}}
\, \cos ({\bf K} \cdot {\bf r}) &  e^{\omega\,t -  \frac{{\bf k \cdot r}}{2}} \, \sin ({\bf K} \cdot {\bf r}) \cr 
 e^{- \frac{\omega\,t}{2}+ {\bf k \cdot r }} \, \cos \Omega \, t & 
e^{- \frac{\omega\,t}{2}- \frac{\bf k \cdot r}{2} } \, \cos (\Omega \, t - {\bf K} \cdot {\bf r}) &
e^{- \frac{\omega\,t}{2} - \frac{{\bf k \cdot r}}{2}} \, \cos (\Omega \, t  - {\bf K} \cdot {\bf r}) \cr
e^{- \frac{\omega\,t}{2} + {\bf k \cdot r}} \, \sin \Omega \, t & 
e^{- \frac{\omega\,t}{2}- \frac{ {\bf k \cdot r}}{2}} \, \sin (\Omega \, t - {\bf K} \cdot {\bf r}) & 
e^{- \frac{\omega\,t}{2} -\frac{{\bf k \cdot r}}{2}} \, \sin (\Omega \, t - {\bf K} \cdot {\bf r}) 
\end{pmatrix}, $$ }
where $\Omega=\frac{\sqrt{3}}{2} \, \omega$ and ${\bf K}=\frac{\sqrt{3}}{2}{\bf k}$; 
the same can be done with the conjugate solutions (with $- \omega$ instead of $\omega$). A similar matrix,
of course, can be produced for the alternative negative $\omega$ choice.

The functions displayed in the matrix do not represent a wave; however, one can produce a propagating
solution by forming certain cubic combinations, e.g. 
{\small
$$e^{\omega\,t + {\bf k \cdot r}} \, e^{- \frac{\omega\,t}{2} - \frac{{\bf k \cdot r}}{2}} 
\, \cos (\Omega \, t - {\bf K} \cdot {\bf r})  \,
e^{- \frac{\omega\,t}{2} - \frac{{\bf k \cdot r}}{2}} \, \sin (\Omega \, t - {\bf K} \cdot {\bf r} ) = 
\frac{1}{2} \, \sin ( 2  \Omega \, t - 2 {\bf K} \cdot {\bf r}). $$}
What we need now is a multiplication scheme that would define triple products of non-propagating solutions yielding
propagating ones, like in the example given above, but under the condition that the factors belong to three distinct
subsets (which can be later on identified as ``colors"). 

Before we proceed farther, let us remind that the set of six independent functions is expected to generate
the most general solution of our sixth-order differential equation. Therefore, among the nine functions displayed
in the above matrices, as well as in the real basis, three are superfluous.
Indeed, the determinants of the two complex matrices of solutions, as well as that of the real
matrix, identically vanish. Their lower $2 \times 2$ minors are also zero, which
confirms the idea that only {\it six} out of nine functions are independent. In principle, we could pick up any
six functions, but for symmetry reasons we shall remove the diagonal ones. The remaining six functions are displayed 
in the truncated matrix:
$$
\begin{pmatrix}   0 &   e^{\omega\,t - \frac{{\bf k \cdot r}}{2}}
\, \cos ({\bf K} \cdot {\bf r}) &  e^{\omega\,t - \frac{{\bf k \cdot r}}{2}} \, \sin ({\bf K} \cdot {\bf r}) \cr 
 e^{- \frac{\omega\,t}{2}+{\bf k \cdot r }} \, \cos \Omega \, t & 0 &
e^{- \frac{\omega\,t}{2}- \frac{{\bf k \cdot r}}{2}} \, \cos \; u \cr
e^{- \frac{\omega\,t}{2} +{\bf k \cdot r}} \, \sin \Omega \, t & 
e^{- \frac{\omega\,t}{2}- \frac{ {\bf k \cdot r}}{2}} \, \sin \; u & 0 
\end{pmatrix}, $$

where $u = \Omega \, t  - {\bf K} \cdot {\bf r}$.

In what follows, we shall choose
the Cartesian system of space coordinates with its $x$-axis aligned with the vector  ${\bf k}$, so that in
all the six remaining functions displayed in the real matrix we can replace the scalar product ${\bf k} \cdot {\bf r}$ by $kx$,
and ${\bf K}\cdot {\bf r}$ by $Kx$, with $K = \frac{\sqrt{3}}{2} \, k$.

With this in mind, let us display the six independent solutions in the following two groups of three:
$$ F_1 = e^{- \frac{\omega t}{2} + kx} \; \sin \Omega t, \; \; \; \; \; \; \;
 F_2 =e^{- \frac{\omega t}{2} + kx} \; \cos \Omega t, $$
$$G_1 = e^{\omega t - \frac{kx}{2}} \; \sin Kx, \; \; \; \; \; \; \; 
G_2 = e^{\omega t - \frac{kx}{2}} \; \cos Kx, $$
\begin{equation}
H_1 = e^{- \frac{\omega t}{2} - \frac{kx}{2}} \; \sin (\Omega t - Kx), \; \; \; \; \; 
H_2 = e^{- \frac{\omega t}{2} - \frac{kx}{2}} \; \cos (\Omega t - Kx).
%\label{FGHbis}
\end{equation} 
Neither of the six functions above can represent a freely propagating wave: even the last two functions,
$H_1$ and $H_2$ contain, besides the running sinusoidal waves, the real exponentials which have a damping effect.
(The wave cannot penetrate distances greater than a few wavelengths, and can last only for times comparable
with few oscillations). 

However, we shall show that certain {\it cubic} expressions can represent
a freely propagating wave, without any damping factors. Taking a closer look at the six
solutions displayed above, we see that the only way to get rid of the real exponents 
present in all those functions, but different damping factors, is to form cubic expressions
constructed with three functions labelled with three {\it different} letters. 

Here is the exhaustive list of {\it eight} admissible cubic combinations: 
$$F_1 \; G_1 \; H_1, \; \; \;  F_2 \; G_1 \; H_1; \; \; \; F_1 \; G_1 \; H_2, \; \; \; F_2 \; G_1 \; H_2;$$
$$F_1 \; G_2 \; H_1, \; \; \; F_2 \; G_2 \; H_1; \; \; \; F_1 \; G_2 \; H_2, \; \; \;  F_2 \; G_2 \; H_2;$$
But these expressions still contain, besides running waves with double frequency $2 \Omega$,
undesirable functions like $\sin \Omega t$ or $\cos K x$. To take an example, we have
$$F_1 \, G_2 \, H_2 = \sin \Omega t \; \cos K x \; \cos (\Omega t - K x) =$$
$$\frac{1}{2} \; \left[ \sin (\Omega t + Kx) + \sin (\Omega t - K x) \right] \; \cos (\Omega t - Kx)=$$
$$\frac{1}{4} \; \sin (2 \Omega t - 2 K x) + \frac{1}{4} \; \sin (2 \Omega t) 
+ \frac{1}{4} \; \sin (2 K x).$$

We omit to give all explicit expressions, in terms of the trigonometric functions, of the eight independent cubic combinations
displayed above, but  we give the final result, showing that there
are only {\it two} combinations of cubic products of solutions of the generalized ternary Dirac equation
that represent running waves, which are the following:
\begin{equation}
F_1 G_2 H_2 + F_1 G_1 H_1 - F_2 G_1 H_2 + F_2 G_2 H_1 = \sin (2 \Omega t - 2 K x),
\label{running1}
\end{equation}
\begin{equation}
F_2 G_2 H_2 + F_2 G_1 H_1 + F_1  G_1 H_2 - F_1 G_2 H_1 = \cos (2 \Omega t - 2 K x).
\label{running2}
\end{equation}
The symmetry of these expressions appears better when grouped as follows:
\begin{equation}
F_1 (G_2 H_2 + G_1 H_1) + F_2 (G_2 H_1 - G_1 H_2) = \sin (2 \Omega t - 2 K x),
\label{running1b}
\end{equation}
\begin{equation}
F_2 ( G_2 H_2 + G_1 H_1) + F_1 ( G_1 H_2 - G_2 H_1) = \cos (2 \Omega t - 2 K x).
\label{running2b}
\end{equation}

%The full set of solutions and their cubic combinations is given in the Appendix III.

Two similar running waves are produced by forming corresponding cubic combinations
of negative frequency solutions obtained by substituting $- \omega$ instead of $\omega$ and $- k$ instead of $ k$. 
The four running waves so obtained could represent  freely
propagating Dirac spinor if the dispersion relation relating $\omega$ and $k$ was
the usual quadratic one, but here it is not. So we are still unable to produce
a Dirac particle from cubic combinations of solutions of our sixth-order system, at least with the same masses of three
particles involved. It is possible that removing the mass degeneracy will make possible construction
of propagating composite fermions.
This scheme is in agreement with the no-go theorem stipulating that the only way to combine the Lorentz symmetry with internal symmetries
is a trivial direct product of groups (as shown in \cite{Raifeartaigh}, \cite{Colemandula})
%It would be interesting to try to solve ternary Dirac equation with some special form of gauge potentials fixed in
%advance, describing the field created by two quarks acting on the third one. 

\section{Propagators}

Let us introduce the Fourier transform of a real function of one variable, and the inverse Fourier transform as follows \cite{Bremerman}:

\begin{equation}
{\hat{f}} (k) =  \, \int_{- \infty}^{\infty} f(x) \; e^{i k x} \; dx, \; \; \; \; 
f(x) = \frac{1}{2 \pi} \, \int_{- \infty}^{\infty} {\hat{f}} (k) \; e^{- i k x} \; d k.
\label{Fourier}
\end{equation}
In this convention, the constant function $f(x) = 1$ is transformed into the Dirac delta function $\delta (k)$ multiplied by $2 \pi$.

In terms of their Fourier transforms, linear differential operators of any order are represented by corresponding algebraical expressions
multiplying the Fourier transform of the unknown function. The Fourier transform of the Green function is then given by the
inverse of this expression, for example, the Fourier transform of the Green function of the Klein-Gordon operator is defined as
$$ \hat{G} (k^{\mu}) = \frac{1}{k_0^2 - {\bf k}^2 - mu^2 },$$
(with $\mu = \frac{mc}{\hbar}$). The Fourier transform of Green's function for the Dirac equation is a $4 \times 4$ matrix:
$$ {\hat{D}} (k^{\mu}) =  \frac{\gamma^{\mu} k_{\mu} + m \; {\mbox{l\hspace{-0.55em}1}}_4 }{-k_0^2 + {\bf k}^2 - m^2},$$
because quite obviously one has
$$ (\gamma^{\mu} k_{\mu} + m \; {\mbox{l\hspace{-0.55em}1}}_4) ( \gamma^{\mu} k_{\mu} - m \; {\mbox{l\hspace{-0.55em}1}}_4 ) = 
 {-k_0^2 + {\bf k}^2 - m^2} \; {\mbox{l\hspace{-0.55em}1}}_4.$$
The ternary generalization of Dirac's equation being written in the most compact form as in (\ref{TernDirac0}), in terms of Fourier
transforms it becomes
\begin{equation}
\left( \Gamma^{\mu} \, k_{\mu} - m \; {\mbox{l\hspace{-0.55em}1}}_{12} \right) \; {\hat{\Psi}} (k) = 0.
\label{Dirterfour}
\end{equation}
TYhe sixth power of the matrix $\Gamma^{\mu} k_{\mu}$ is diagonal and proportional to $m^6$, so that we have
\begin{equation}
 \left( \Gamma^{\mu} k_{\mu} \right)^6 - m^6  \; {\mbox{l\hspace{-0.55em}1}}_{12}   = 
\left( k_0^6 - \mid {\bf k} \mid^6 - m^6 \right) \;  {\mbox{l\hspace{-0.55em}1}}_{12} =  0.
\label{Gammasixem}
\end{equation}
Now we have to find the inverse of the matrix  $\left( \Gamma^{\mu} \, k_{\mu} - m \; {\mbox{l\hspace{-0.55em}1}}_{12} \right)$. 
To this effect, let us note that the sixth-order expression on the left-hand side in (\ref{Gammasixem}) can be factorized as follows:
\begin{equation}
\left( \Gamma^{\mu} k_{\mu} \right)^6 - m^6 = \left( \left(\Gamma^{\mu} k_{\mu} \right)^2 - m^2 \right) \;
\left(  \left( \Gamma^{\mu} k_{\mu} \right)^2 - j \; m^2 \right) \; \left( \left( \Gamma^{\mu} k_{\mu} \right)^2 - j^2 \; m^2 \right).
\label{Gammafact3}
\end{equation}
The first factor is in turn the product of two linear expressions, one of which is the ternary Dirac operator:
\begin{equation}
\left( \Gamma^{\mu} k_{\mu} \right)^6 - m^6 = 
\left( \Gamma^{\mu} k_{\mu} - m \right) \;\left( \Gamma^{\mu} k_{\mu} + m \right) \;
\left(  \left( \Gamma^{\mu} k_{\mu} \right)^2 - j \; m^2 \right) \; \left( \left( \Gamma^{\mu} k_{\mu} \right)^2 - j^2 \; m^2 \right).
\label{Gammafact4}
\end{equation}
Therefore the inverse of the Fourier transform of the ternary Dirac operator is given by the following matrix:
\begin{equation}
\left[ \left( \Gamma^{\mu} k_{\mu} \right)^6 - m^6 \right]^{-1} = 
\frac{\left( \Gamma^{\mu} k_{\mu} + m \right) \;
\left(  \left( \Gamma^{\mu} k_{\mu} \right)^2 - j \; m^2 \right) \; 
\left( \left( \Gamma^{\mu} k_{\mu} \right)^2 - j^2 \; m^2 \right)}{\left( k_0^6 - \mid {\bf k} \mid^6 - m^6 \right)}.
\label{Dirac3inverse}
\end{equation}
It takes almost no effort to prove that the numerator can be given a more symmetric form. Taking into account that
$$\left(  \left( \Gamma^{\mu} k_{\mu} \right)^2 - j \; m^2 \right) \left( \left( \Gamma^{\mu} k_{\mu} \right)^2 - j^2 \; m^2 \right) =
\left( \Gamma^{\mu} k_{\mu} \right)^4 + m^2 \; \left( \Gamma^{\mu} k_{\mu} \right)^2 + m^4,$$
we find that
$$\left( \Gamma^{\mu} k_{\mu} + m \right) 
\left(  \left( \Gamma^{\mu} k_{\mu} \right)^2 - j \; m^2 \right)  
\left( \left( \Gamma^{\mu} k_{\mu} \right)^2 - j^2 \; m^2 \right) = $$
$$ \left( \Gamma^{\mu} k_{\mu} \right)^5 + 
m \;  \left( \Gamma^{\mu} k_{\mu} \right)^4 + m^2 \;  \left( \Gamma^{\mu} k_{\mu} \right)^3 + 
m^3 \;  \left( \Gamma^{\mu} k_{\mu} \right)^2 + m^4 \;  \left( \Gamma^{\mu} k_{\mu} \right) + m^5,$$
so that the final expression can be written in a concise form as
\begin{equation}
\left[ \left( \Gamma^{\mu} k_{\mu} \right)^6 - m^6 \right]^{-1} = 
\frac{{\displaystyle \sum_{s=0}^5} m^s \; \left( \Gamma^{\mu} k_{\mu} \right)^{(5-s)}}{\left( k_0^6 - \mid {\bf k} \mid^6 - m^6 \right)}.
\label{Dirac3invsymb}
\end{equation}

In the massless case, the operator equation whose Green's function we want to evaluate, reduces to
$$ \left[ \frac{1}{c^6} \frac{\partial^6}{\partial t^6} - \left( \frac{\partial^2}{\partial x^2} + \frac{\partial^2}{\partial y^2}
+ \frac{\partial^2}{\partial z^2} \right)^3 \right] \; G (t, {\bf r}) = \delta^4 (x) = \delta(ct) \delta(x) \delta(y) \delta(z).$$
Using the Fourier transformation method, we can write:
\begin{equation} 
\left[ \frac{\omega^6}{c^6} - \mid {\bf k} \mid^6 \right] \; {\hat{G}} (k_0, {\bf k}) = 1, \; \; \; \; {\rm where} \; \; k_0 = \frac{\omega}{c},
\label{FourierGreen}
\end{equation}
from which we get 
\begin{equation}
{\hat{G}} (k_0, {\bf k} ) =  \frac{1}{k_0^6 - \mid {\bf k} \mid^6 } + \Phi (k_0, {\bf k} ),
\label{FGreen2}
\end{equation}
where $\Phi (k_0, {\bf k} )$ is a solution of the homogeneous equation, 
\begin{equation}
\left[ k_0^6 - \mid {\bf k} \mid^6 \right] \; \Phi (k_0, {\bf k} ) = 0 \; \; \; {\rightarrow} \; \; \; \Phi (k_0, {\bf k} ) = \delta (k_0^6 - \mid {\bf k} \mid^6 ).
\end{equation} 
The sixth-order polynomial $k_0^6 - \mid {\bf k} \mid^6 $ can be split into the product of three second-order factors as follows:
\begin{equation}
k_0^6 - \mid {\bf k} \mid^6 = (k_0^2 - \mid {\bf k} \mid^2) \; (k_0^2 - j \mid {\bf k} \mid^2) \; (k_0^2 - j^2  \mid {\bf k} \mid^2),
\label{kthree}
\end{equation}
each of which being a product of two linear expressions with opposite signs of $\mid {\bf k} \mid $:
$$ (k_0^2 - \mid {\bf k} \mid^2) = (k_0 + \mid {\bf k} \mid ) \,(k_0 - \mid {\bf k} \mid ), $$ 
$$ (k_0^2 -  j \, \mid {\bf k} \mid^2) = (k_0 + j^2 \; \mid {\bf k} \mid ) \,(k_0 - j^2 \; \mid {\bf k} \mid ), $$ 
$$ (k_0^2 - j^2 \,\mid {\bf k} \mid^2) = (k_0 + j \; \mid {\bf k} \mid ) \,(k_0 - j \; \mid {\bf k} \mid ), $$ 
so that the sixth-order expression appearing in (\ref{FourierGreen}) can be decomposed into a product of six linear terms.
Let us represent the inverse of this expression appearing in (\ref{FGreen2}) as a sum of three fractions with second-order
expressions in their denominators:
\begin{equation}
\frac{1}{k_0^6 - \mid {\bf k} \mid^6 } = \frac{1}{3 \mid {\bf k} \mid^4} \;  \left[ \frac{1}{k_0^2 - \mid {\bf k} \mid^2 }
 + \frac{j}{k_0^2 - j \, \mid {\bf k} \mid^2 } + \frac{j^2}{k_0^2 - j^2 \; \mid {\bf k} \mid^2 } \right],
\label{inversesix}
\end{equation}
which is to be compared with the usual Fourier inverse of the d'Alembert operator:
\begin{equation}
\frac{1}{k_0^2 - \mid {\bf k} \mid^2 } = \frac{1}{2 \mid {\bf k} \mid^2} \;  \left[ \frac{1}{k_0 - \mid {\bf k} \mid }
 - \frac{1}{k_0 +  \, \mid {\bf k} \mid } \right]
\label{inversetwo}
\end{equation}
The difference in the order of the equation leads to the difference in the algebraic structure of the polynomial representig
the equation for the Fourier transform. Its inverse displays not just two, but as much as {\it six} simple poles displayed in the following figure: 
\begin{figure}[hbt]
\centering 
\includegraphics[width=6cm, height=5.2cm]{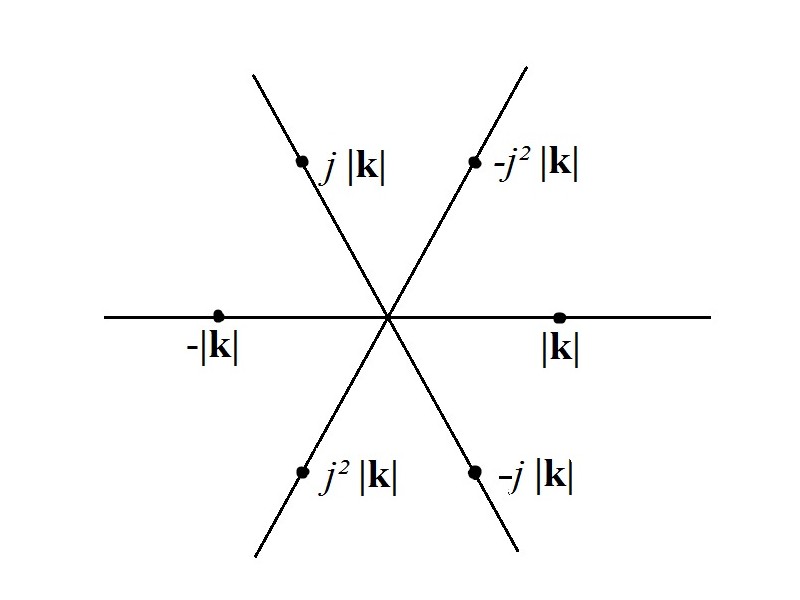}
\caption{{\small The six simple poles of the integral representation of zero-mass propagator
of the sixth-order equation}}
\label{fig:Sixpoints}
\end{figure}
In the case of the usual d'Alembertian several different Green's functions can be obtained by taking the inverse Fourier transform of the (\ref{inversetwo}),
The most widely used is the {\it retarded} Green's function, proportional to the well known expression
\begin{equation}
G_{ret} (x^{\mu}) = Y(ct) \; \frac{\delta (ct - kr)}{4 \pi r}
\label{Greenret}
\end{equation}
where $Y(ct)$ is Heaviside's function.

In our case the Fourier transform of the Green function we are looking for is a product of three factors, namely
 we have to do with a product of three factors, namely:
\begin{equation}
{\hat{G}}_1 =   \left[ \frac{1}{k_0^2 - \mid {\bf k} \mid^2 } \right] \; \; \; 
{\hat{G}}_2 =   \left[ \frac{j}{k_0^2 - j \, \mid {\bf k} \mid^2 } \right] \; \; {\rm and} \; \; \; 
{\hat{G}}_3 =   \left[\frac{j^2}{k_0^2 - j^2 \; \mid {\bf k} \mid^2 } \right],
\label{inversesixproduct}
\end{equation}
%It could be decomposed into a sum of three terms:
%\begin{equation}
%{\hat{G}} =  \frac{1}{3 \mid {\bf k} \mid^4} \;  \left[ \frac{1}{k_0^2 - \mid {\bf k} \mid^2 } \right]
%+  \frac{1}{3 \mid {\bf k} \mid^4} \;  \left[ \frac{j}{k_0^2 - j \, \mid {\bf k} \mid^2 } \right]
%+ \frac{1}{3 \mid {\bf k} \mid^4} \;  \left[\frac{j^2}{k_0^2 - j^2 \; \mid {\bf k} \mid^2 } \right],
%\label{inversesixsum}
%\end{equation}
%each of them presenting a strong infrared divergence due to the factor $\frac{1}{\mid {\bf k} \mid^4}$, thus making quite difficult
%the inverse Fourier integration necessary for obtaining the original Green function:

\begin{equation}
G (x^{\mu}) = \frac{1}{16 \pi^4} \; \int_0^{2 \pi} d \varphi  \int_0^{\pi} \sin \theta d \theta \int_0^{\infty} {\bf k}^2 d \mid {\bf k} \mid 
\int_{- \infty}^{\infty} d k_0  e^{-i (k_0 c t - \mid {\bf k} \mid r \cos \theta)}  \;  {\hat{G}} (k^{\mu}),
\label{Fourier4}
\end{equation}
With ${\hat{G}} (k^{\mu})$ given by the expression (\ref{inversesix} or the product ${\hat{G}}_1 \; {\hat{G}}_2 \; {\hat{G}}_3 $ (\ref{inversesixproduct}).
Qualitative picture is much easier to obtain i we use the latter form, because the inverse Fourier transform of a product is equal to the convolution
of inverse Fourier transforms of each factor. And the inverse Fourier transforms of each of the three factors defined in (\ref{inversesixproduct}) can be  
found using the standard integration procedure described in textbooks on Fourier transforms or in standard textbooks on electrodynamics
(\cite{LandauLifshitz}, \cite{Bremerman}). 

Each of its three factors contains an inverse of quadratic expression
resembling the usual d'Alembertian, with $\mid {\bf k}^2 \mid$ appearing with factors $1, \; j$ and $j^2$.

In what follows, we shall write $k$ instead of $\mid {\bf k} \mid$ when there is no risk of ambiguity. 
Supposing that $G (k^{\mu})$ is spherically symmetric, the integration over $d \varphi$ gives just the factor $2 \pi$. Next, we can perform
integration over  $ d \theta$, factorizing the only term depending on $\theta$, which is $e^{i k r \cos \theta}$. This integral gives
\begin{equation}
\int_0^{\pi} \; e^{i k r \cos \theta} \; \sin \theta d \theta = \int_{-1}^1 \; e^{i k r u} \, du = \frac{2 \, \sin kr}{kr}.
\label{sinkr}
\end{equation}
What remains now is the integration over $d k$ and $d k_0$. As usual, the integral over $dk_0$ is taken first, and evaluated
by extention to the complex domain. The first factor $G_1$ has two poles on the real line, $k_0 = \pm \mid {\bf k} \mid$, and is evaluated
as a principal value. The final result is the well known Green's function of the d'Alembertian, \ref{Greenret}. The remaining two factors
are integrated over $d k_0$ even more easily, because their poles are found off the real axis, as shown in the following figure:
\begin{figure}[hbt]
\centering
\includegraphics[width=6.6cm, height=4.6cm]{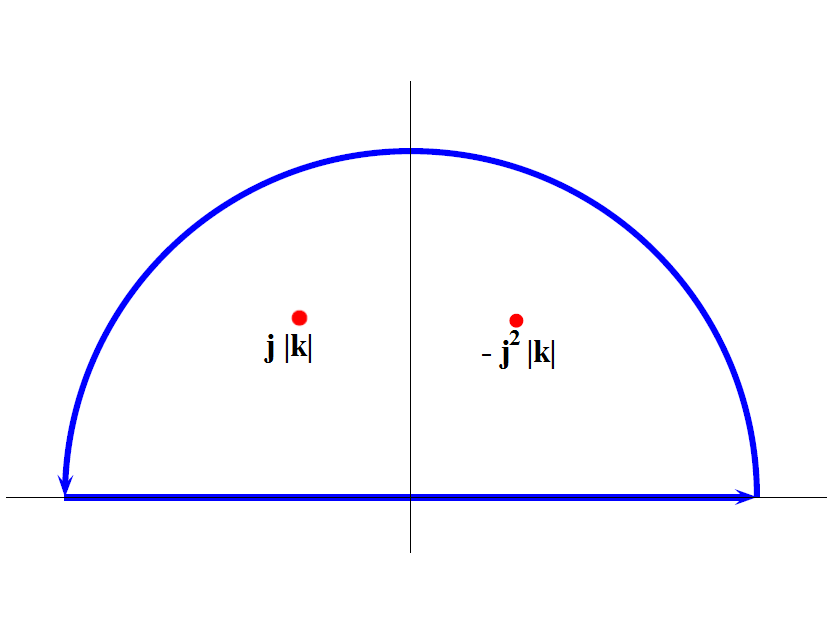}
\hskip 0.2cm
\includegraphics[width=6.6cm, height=4.6cm]{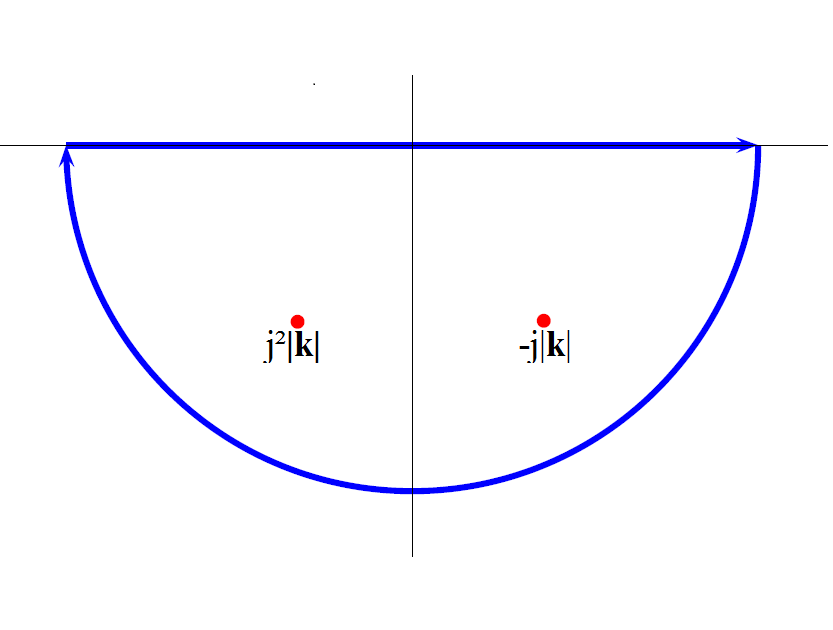}
\caption{{\small Left: The upper contour, containing the poles at $j \mid {\bf k} \mid$ and $- j^2 \; \mid {\bf k} \mid$, for $t < 0$; 
Right: The lower contour, containing the poles at $ -j \mid {\bf k} \mid$ and $j^2 \; \mid {\bf k} \mid$, for $t > 0$ . }}
\label{fig:Contours}
\end{figure}
The resulting integrals yield the following expressions:
\begin{equation}
- 2 \pi i \; Y(t) \; \left[j \; e^{ij k ct} - j^2 \; e^{i j^2 k ct} \right] \; \; \; \; \;  {\rm and} \; \; \; \; \;  
2 \pi i \; Y(- t) \; \left[ j^2 \;e^{ij^2 k ct} - j \; e^{- i j k ct} \right] 
\label{TwoYj}
\end{equation}
Substituting explicit expressions for all complex numbers appearing in these expressions, we get two real functions:
$$ 2 \pi \; Y(t)  \; e^{- \frac{\sqrt{3}}{2} k ct} \; \left[ \sqrt{3} \cos \frac{k }{2} c t + \sin \frac{k}{2} ct \right],
\; \; \; \;  2 \pi \; Y(- t)  \; e^{\frac{\sqrt{3}}{2} k ct} \; \left[ \sqrt{3} \cos \frac{k }{2} c t - \sin \frac{k}{2} ct \right],$$
Both expressions contain the damping factor $e^{\frac{\sqrt{3}}{2} k ct}$ which is absent in the first contribution proportional 
to the usual d'Alembertian. But as all three components mix together, all will acquire these damping factors
and fade away very quickly.
These expressions multiply the Fourier transforms of each of the three $k$-dependent parts of Fourier images of Green's function,
 $G_2$ and $G_3$, while the first one, $G_1$ similar to the usual d'Alembertian, has to be multiplied by $2 \pi \; Y(t) \; \sin (kct)$. 
Before performing the last integration over $dk$, they should be multiplied by the factor $\frac{2 \pi \; \sin (kr)}{kr}$.

The final results for each of the factors become, as could be expected, as follows:
$$ G_1  = Y(ct) \; \frac{\delta (k ct - k r)}{r}, \; \; \; G_2  = Y(ct) \; \frac{ \delta (kct - j^2 \; k r)}{r},
\; \; \; G_3  = Y(ct) \; \frac{ \delta (kct - j \; k r)}{r}. $$
and the Green's function of the sixth-order massless operator can be obtained by the convolution of the three functions:
$$G (x^{\mu}) = G_1 * G_2 * G_3.$$
The Dirac $\delta$-functions vanish everywhere except for the light cone $kct - kr = 0$ in $G_1$, and for complex-valued
wave vectors $k$ such that for $G_2 (x^{\mu})$ $k$ must be proportional to $j$, and in the case 
of $G_3 (x^{\mu})$ $k$ must be aligned along the $j^2$ axis. 

After second quantization, these solutions can be implemented as operators with well-defined commutation properties. In order
to reproduce the existing stable configurations of quarks of the first generation, $uud$ and $udd$, a generalized Pauli's exclusion principle
based on the $Z_3$ symmetry should replace the usual $Z_2$ symmetric exclusion principle, as proposed in 
%\cite{Kerner1983}, \cite{Kerner{1991}, \cite{Kerner1992}, 
\cite{Kerner2013}, \cite{Kerner2014} and \cite{Kerner2017}.
\vskip 0.4cm
\indent
\hskip 0.5cm
{\bf {\large {Acknowledgements}}}
\vskip 0.2cm
\indent
\hskip 0.5cm
I am greatly indebted to Michel Dubois-Violette, Viktor Abramov and Karol Penson for many discussions and constructive criticism.
I would like to express my sincere thanks to Jan-Willem van Holten, J\"urg Fr\"olich, Yuri Dokshitser, Paul Sorba and Reinald Flume for important 
discussions, suggestions and remarks. Thanks are due to Dr. Katarzyna G\'orska for her help with symbolic calculus.

\newpage

\indent
\hskip 0.5cm
{\large {\bf References}}


\vskip 0.3cm

\begin{thebibliography}{90}

\bibitem{GellMannNeeman} M. Gell-Mann, Y. Ne'eman, The Eightfold Way, Benjamin, New York (1964)

\bibitem{Lipkin} H.J. Lipkin, {\it Frontiers of the Quark Model}, Weizmann
Inst. pr. WIS-87-47-PH (1987)

\bibitem{Okubo} S. Okubo, Journ. of Math. Physics, {\bf 34}, 3273;
{\it ibid ,} 3292  (1993)

\bibitem{Scherk1975} J. Scherk {\it Rev. Mod. Phys}. {\bf 47}, 123 (1975)

\bibitem{WessMadore} Madore J, Schraml S, P. Schupp P, Wess J, {\it The European Physical Journal C}, {\bf 16} (1), pp 161-167 (2000)

\bibitem{AKL2015} Abramov V, Kerner R, Liivaapuu O 2015 - arXiv preprint arXiv:1512.02106, 2015 - arxiv.org 


%\bibitem{WessZumino} Wess J and Zumino B  {\it Nuclear Physics B} {\bf 70}, pp.39-50 (1974).

%\bibitem{SternGerlach} W. von Gerlach, O. Stern, {\it Zeitschrift f\"ur Physik}, {\bf 8}, p.110 (1921);
%{\it ibid}: {\bf 9}, p. 349 (1922).

\bibitem{UhlenbeckGoudsmit} S. Goudsmit and G.E. Uhlenbeck, {\it Physica} {\bf 6}, p. 273 (1926); see also
G.E. Uhlenbeck and S. Goudsmit, {\it Nature} {\bf 117}, p. 264 (1926).

\bibitem{Pauli} W. Pauli, {\it Zeitschrift f\"ur Physik}, {\bf 43}, p. 601-623  (1927)

\bibitem{Dirac} P.A.M. Dirac, {\it Proc. Royal. Soc. (London)}, A {\bf 117}, p. 610-624  (1928);
{\it ibid} A {\bf 118}, p. 351-361 (1928).

\bibitem{Sogami2013} Sogami I S,  {\it Progress of Theoretical and Experimental Physics}, Volume 2013, Issue 12, (2013), 
123B02, https://doi.org/10.1093/ptep/ptt103

\bibitem{Sylvester} Sylvester, J. J {\it Johns Hopkins University Circulars} I: 241-242; ibid II (1883) 46; ibid III pp 7-9 (1884). 

\bibitem{VKac1994} V. Kac, Infinite dimensional Lie algebras, Cambridge University Press, 1994.

\bibitem{LiWei} Yu L-W, Ge M-L, {\it Scientific Reports},  6:21497, DOI: 10.1038/srep21497 (2016)

\bibitem{MDVRKJM1} Dubois-Violette M, Kerner R and Madore J, {\it Journ. Math. Phys.}, {\bf 31} (2), 316-323, (1990)

\bibitem{MDVRKJM2} Dubois-Violette M, Kerner R and Madore J, {\it Journ. Math. Phys.}, {\bf 31} (2), 323-331, (1990)

\bibitem{Shadow} Dubois-Violette M, Madore J, Kerner R, {\it Journ. Math. Phys.}, {\bf 39} (2), 730, (1998)

\bibitem{Kerner1983} Kerner R 1983 {\it Communications in mathematical physics} {\bf 91} (2), pp. 213-234

\bibitem{Raifeartaigh} O'Raifeartaigh L., {\it Physical Review} 139, B1052 (1965)

\bibitem{Colemandula} Coleman S., Mandula J. {\it Physical Review} 159, p. 1251 (1967)

\bibitem{LandauLifshitz} Landau L.D, Lifshitz E.M, {\it The Classical Theory of Fields}, Third Revised edition, Pergamon Press (1971)  

\bibitem{Bremerman} Bremerman H., {Distributions, Complex Variables and Fourier Transforms}, Addison-Wesley, Mass., USA (1965)

%\bibitem{Kerner1983} Kerner R 1983 {\it Communications in mathematical physics} {\bf 91} (2), pp. 213-234

\bibitem{Kerner1991} Kerner R 1991 {\it Comptes Rendus Acad. Sci. Paris.}  {\bf 10} pp. 1237-1240

\bibitem{Kerner1992} Kerner R {\it Journal of Mathematical Physics}, {\bf 33} (1) pp.403-4011 (1992)

\bibitem{Kerner2013} Kerner R in {\it Symmetrties and Groups in Contemporary Physics}, World Scientific, 
eds. Chengming Bai, J.-P.Gazeau, Mo-Lin Ge), pp. 283-288 (2013) 

\bibitem{Kerner2014}  Kerner R {\it Algebra, Geometry and Mathematical Physics}, in Springer Proceedings series, Math.Stat., ed. A. Makhlouf and E. Paal, 
{\bf 85} pp. 617-637 (2014)

\bibitem{Kerner2017} Kerner R {\it Physics of Atomic Nuclei} {\bf 80} (3), pp. 529-541 (2017) 

\end{thebibliography}
\end{document}